\documentclass[journal]{IEEEtran}
\ifCLASSOPTIONcompsoc
  \usepackage[nocompress]{cite}
\else
  \usepackage{cite}
\fi

\ifCLASSOPTIONcompsoc \usepackage[caption=false,font=normalsize,labelfon
t=sf,textfont=sf]{subfig}
\else
\usepackage[caption=false,font=footnotesize]{subfi g}
\fi
\ifCLASSINFOpdf
  \usepackage[pdftex]{graphicx}
\else
\fi
\usepackage{multirow}

\usepackage{tikz}
\usetikzlibrary{arrows,positioning} 

\usepackage{amsmath}
\usepackage{amsthm}
\theoremstyle{definition}
\newtheorem{definition}{Definition}

\usepackage{array}
\def\negspace{\mkern-10mu}
\def\negdoublespace{\mkern-20mu}

\def\myjournaltitle{On Reducing IoT Service Delay via Fog Offloading}

\usepackage[pscoord]{eso-pic}
\newcommand{\placetextbox}[3]{
\setbox0=\hbox{#3}
\AddToShipoutPictureFG*{ \put(\LenToUnit{#1\paperwidth},\LenToUnit{#2\paperheight}){\vtop{{\null}\makebox[0pt][c]{#3}}}
}
}
\placetextbox{.5}{0.055}{\scriptsize{Copyright (c) 2018 IEEE. Personal use is permitted. For any other purposes, permission must be obtained from the IEEE by emailing pubs-permissions@ieee.org.}}

\begin{document}
\title{\myjournaltitle}


\author{Ashkan~Yousefpour,~\IEEEmembership{Student Member,~IEEE,}
	Genya~Ishigaki,~\IEEEmembership{Student Member,~IEEE,}
	Riti~Gour,
	and~Jason~P.~Jue,~\IEEEmembership{Senior~Member,~IEEE}% <-this % stops a space
	\thanks{The authors are with the Advanced Network Research Lab, Department
		of Computer Science, The University of Texas at Dallas, Richardson, TX, 75080 USA. Email: \{ashkan, gishigaki, riti.gour, jjue\}@utdallas.edu}% <-this % stops a space
	\thanks{An earlier version of this paper appeared in IEEE EDGE 2017, Honolulu, Hawaii \cite{ashkan-fog-delay}.}}


% compsoc conferences compared to traditional conferences), use this
%Telephone: (800) 555--1212, Fax: (888) 555--1212}


\markboth{}%
{Yousefpour \MakeLowercase{\textit{et al.}}: \myjournaltitle}

\maketitle

\begin{abstract}
With the Internet of Things (IoT) becoming a major component of our daily life, understanding how to improve the quality of service (QoS) for IoT applications through fog computing is becoming an important problem. In this paper, we introduce a general framework for IoT-fog-cloud applications, and propose a delay-minimizing collaboration and offloading policy for fog-capable devices that aims to reduce the service delay for IoT applications. We then develop an analytical model to evaluate our policy and show how the proposed framework helps to reduce IoT service delay. 
\end{abstract}



\IEEEpeerreviewmaketitle
% \fi
\section{Introduction}
\IEEEPARstart{T}{he} Internet of Things (IoT) is likely to be incorporated into our daily life, in areas such as transportation, healthcare, industrial automation, smart home, smart city, and emergency response. The IoT enables things to see and sense the environment, to make coordinated decisions, and to perform tasks based on these observations \cite{IoTsurvey}. In order to realize the full benefits of the IoT, it will be necessary to provide sufficient networking and computing infrastructure to support low latency and fast response times for IoT applications. Cloud Computing has been the main enabler for IoT applications with its ample storage and processing capacity. Nonetheless, being far from end-users, cloud-supported IoT systems face several challenges including high response time, heavy load on cloud servers and lack of global mobility.

In the era of Big Data, it may be inefficient to send the extraordinarily large amount of data that swarms of IoT devices generate to the cloud, due to the high cost of communication bandwidth, and due to the high redundancy of data (for instance, constant periodic sensor reading). Instead of moving data to the cloud, it may be more efficient to move the applications, storage, and processing closer to the data produced by the IoT. Fog computing is well suited to address this issue by moving the above services closer to where data is produced.

Fog computing is an emerging concept that puts the cloud services closer to the end users (and things) for better QoS \cite{chiang2016fog, Cisco}. Fog is an intelligent layer sitting between cloud and IoT, that brings low latency, location awareness, and wide-spread geographical distribution for the IoT. Inheriting main concepts from cloud computing, fog provides computation, storage, and networking services to end-users, anywhere along the thing-to-cloud continuum, according to OpenFog Consortium. The idea is to serve the requests that demand real-time and low-latency services at the fog, and to send the requests that demand permanent storage or require extensive analysis to the cloud \cite{communicationMagazine,mobility-aware-scheduling}. Due to the countless benefits of fog, the research in this area has been gaining attention, and researchers have recently started to define visions, basic notions, and possible architectures of fog computing \cite{Cisco, network_sdn, chiang2016fog}.
\subsection{Related Work}
The problem of task offloading in fog has recently gained attention from researchers. The authors in \cite{two-datacenter} propose and formulate a cooperative  offloading policy between two fog data centers for load balancing. Similarly, the work in \cite{ring-datacenter} analyzes an offloading policy between multiple fog data centers in a ring topology. In this paper we formulate and propose a general policy for fog nodes to offload IoT requests to other fog nodes or to forward them to the cloud, in order to minimize the IoT service delay. This policy is general, in the sense that the number and topology of fog nodes are arbitrary. There are other similar efforts aimed at minimizing delay via fog computing, which are not directly related to delay-minimizing fog offloading. For instance, the authors in \cite{communicationMagazine} introduce a hybrid architecture that integrates a cloud radio access network and a fog radio access network to handle network traffic. They propose the idea of serving delay-tolerant traffic in the cloud and handling low-latency traffic in the fog. 

A similar problem to task offloading in the fog (fog offloading), is the task or workload assignment problem in the fog, which is assigning tasks or workloads to either fog nodes or cloud servers, while minimizing delay, cost, or energy. In these problems, the set of tasks are given and the problem is modeled as a static optimization problem that determines the assignment of the tasks or workloads. Namely, the authors in \cite{MedicalCPS} study base station association, task distribution, and virtual machine placement in fog-computing-supported medical cyber-physical systems, while minimizing the overall cost and satisfying QoS requirements. Another effort is the work in \cite{power-delay-tradeoff} that addresses power-delay trade-off in cloud-fog computing by workload allocation. More recent work in \cite{Assessment} focuses on theoretical modeling of fog computing architectures, specifically, service delay, power consumption, and cost. Nevertheless, task assignment problems are static in nature and have high complexity. In task assignment problems, all the parameters (of nodes and network) are assumed to be known to an entity that solves the overall problem for the optimal solution, which may not be practical. The fog offloading problem does not have such assumptions.  
	
%The paper has strong contributions, the suggested scheme cannot be generalized to IoT-fog-cloud scenarios, as it is based on the cellular network architecture. re-adjust
%The scheme lacks the cloud entity, and there is no computation offloading capability. 
%This scheme is similar to that of \cite{indexbased} in the sense that it tries to achieve a power-delay trade-off in edge clouds. The authors mathematically formulate the workload allocation problem in fog-cloud computing. 

Recent work in \cite{indexbased} addressed the design of a policy for assigning tasks that are generated by mobile subscribers to edge clouds, to achieve a power-delay trade-off. The problem is first formulated as a static optimization problem. The authors then propose a distributed policy for task assignment that could be implemented efficiently on the edge clouds. However, IoT-to-cloud or fog-to-cloud communication and computation offloading capability are not considered. Moreover, in the distributed task assignment policy, edge clouds broadcast their status continuously to mobile subscribers, which may limit scalability in IoT networks with a large number of edge devices and end users. 

In this paper we study the problem of fog offloading for reducing IoT service delay, which is a fundamentally different problem from the problems discussed above (e.g. \cite{communicationMagazine,MedicalCPS,indexbased,Assessment,power-delay-tradeoff}). The proposed offloading policy is general as opposed to studies \cite{two-datacenter,ring-datacenter}, as it does not place any restrictions on the number or topology of the fog nodes. In the rest of the paper, we discuss our proposed fog framework and offloading policy.
\subsection{Contribution}
In this work, we introduce a common sense general framework to understand, evaluate and model service delay in IoT-fog-cloud application scenarios. We then propose a delay-minimizing offloading policy for fog nodes whose goal is to reduce service delay for the IoT nodes. In contrast to the existing works in the literature, the proposed policy considers IoT-to-cloud and fog-to-cloud interaction, and also employs fog-to-fog communication to reduce the service delay by sharing load. For load sharing, the policy considers not only the queue length but also different request types that have variant processing times. Additionally, our scheme is not limited to any particular architecture (such as a cellular network), and the number, type, or topology of IoT nodes, fog nodes, and cloud servers are not restricted. We also develop an analytical model to evaluate service delay in the IoT-fog-cloud scenarios, and perform extensive simulation studies to support the model and the proposed policy. This paper extends our previous work \cite{ashkan-fog-delay} as we introduce a complete analytical model for evaluating service delay, and we present additional results.
\subsection{Paper Organization}
The rest of this paper is organized as follows. We introduce our IoT-fog-cloud framework in Section \ref{model} and propose the policy for reducing IoT service delay in Section \ref{policy}. We then formally introduce the analytical model for evaluating IoT service delay and its components in Section \ref{analytical}, and explain the numerical results of our experiment in Section \ref{results}. Finally, Section \ref{conclusion} summarizes the paper and provides future directions for this work.  
\section{General IoT-Fog-Cloud Framework} \label{model}
\begin{figure}[!t]
 	\includegraphics[width=\linewidth]{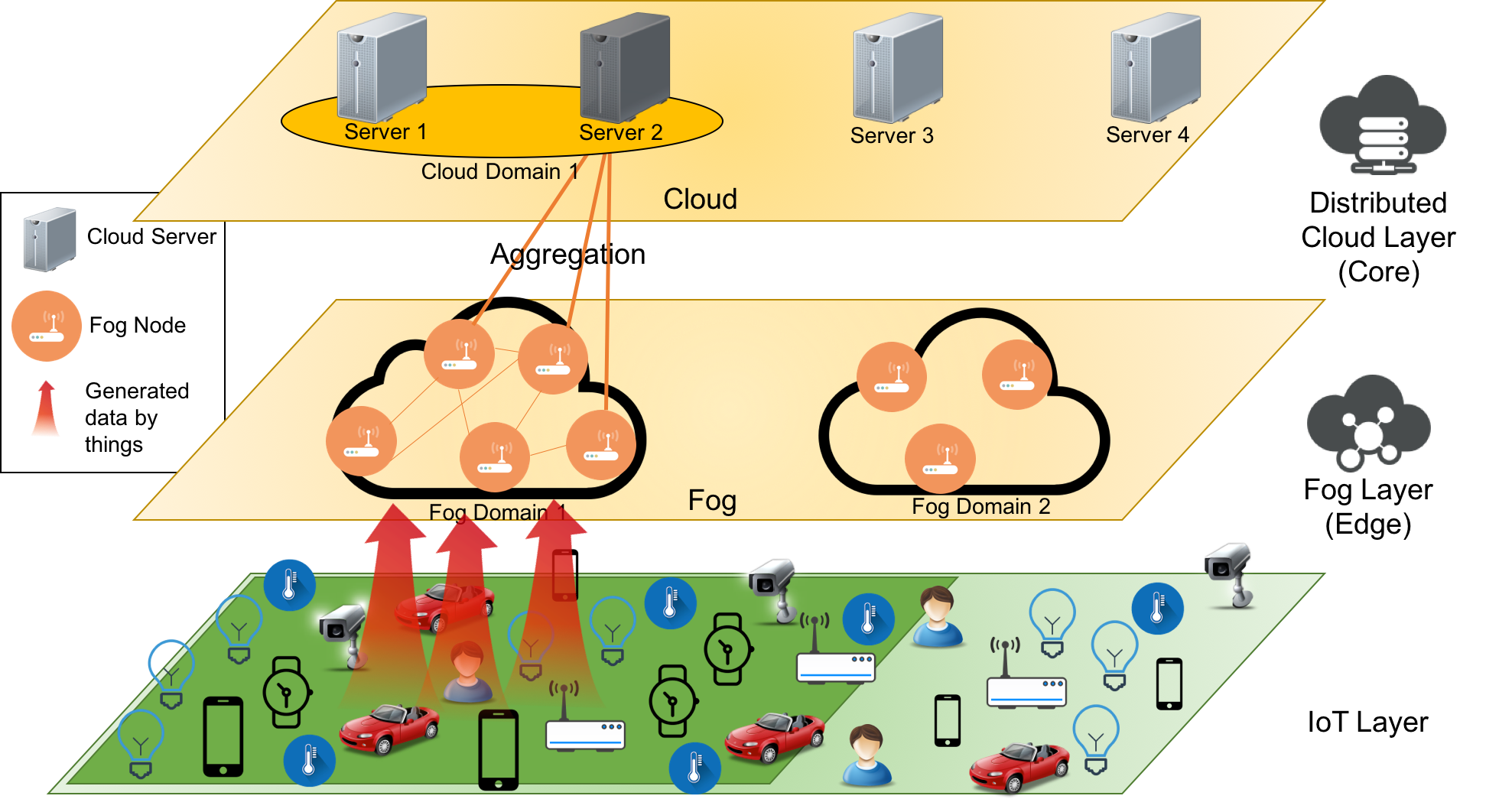}
 	\caption{General framework for IoT-fog-cloud architecture. Each layer is partitioned into domains where a single application is implemented.}
 	\label{fig:arch}
 \end{figure}

Figure \ref{fig:arch} illustrates a general framework for an IoT-fog-cloud architecture that is considered in this work. There are three layers in this architecture: IoT layer, where the IoT devices and end-users are located, fog layer, where fog nodes are placed, and cloud layer, where distributed cloud servers are located. A \textit{cloud server} can be composed of several processing units, such as a rack of physical servers or a server with multiple processing cores. In each layer, nodes are divided into domains where a single IoT-fog-cloud application is implemented. 

For instance, a domain of IoT nodes (in a factory, for instance) is shown in dark green, and they communicate with a domain of fog nodes associated with the application. A domain of IoT nodes could comprise IoT devices in a smart home, temperature sensors in a factory, or soil humidity sensors in a farm  where all the IoT devices in the vicinity are considered to be in a single domain. Normally the fog nodes in one domain are placed in close proximity to each other, for example, in a neighborhood or in levels of a building. Each domain of fog nodes is associated with a set of cloud servers for a single application. 

The basic way in which IoT nodes, fog nodes, and cloud servers operate and interact is as follows. IoT nodes can process requests locally, send it to a fog node, or send it to the cloud; fog nodes can process requests, forward requests to other fog nodes in the same domain, or forward the requests to the cloud; cloud servers process requests and send the response back to the IoT nodes. In this work, the aim is to minimize service delay for IoT devices in the proposed framework based on fog computing. The fog layer lies between IoT devices and the cloud; thus, it can handle a majority of IoT service requests in order to reduce the overall service delay.
\begin{definition}
{\em Service delay} for an IoT node is the time required to serve a request, i.e. the time interval between the moment when an IoT node sends a service request and when it receives the response for that request.
\end{definition}

We first introduce the delay-minimizing policy in the following section, then formulate the IoT service delay to evaluate the policy analytically. 
\section{Fog Node Collaboration Policy} \label{policy}
In this section, we introduce the framework in which fog nodes collaborate with each other to fulfill the requests sent from IoT nodes to the fog layer. If a fog node can accept a request based on its current load, it processes the request; however, when the fog node is busy processing many tasks, it may offload the request to some other fog nodes or to the cloud (this is called {\em load sharing}). The concept of load sharing is well studied in the literature \cite{LS1,LS2}, and we borrow similar concepts for the design of the policy by which fog nodes collaborate. This collaboration is discussed in detail the following subsections. 
\subsection{Modes of Interaction}
We propose two modes of interaction for fog nodes: one mode is centralized, in which a central authority controls the interaction of fog nodes, and the other one is distributed, where fog nodes interact with their neighboring fog nodes using a universal protocol.

In the central mode of interaction, in each domain of fog nodes there is a Central Node that controls the interaction among the fog nodes in that domain. In particular, based on the current condition of their queue, fog nodes in domain $\mathcal{M}$ report their estimated time to process the requests in queue and in service (i.e. estimated waiting time) to the Central Node of domain $\mathcal{M}$.

\begin{definition}
	{\em Estimated waiting time} ($W$) is the sum of the estimated processing delays of the requests in the queue (queueing delay), plus the estimated processing delay of the request under service.
\end{definition}

The Central Node of domain $\mathcal{M}$ announces the estimated waiting time of the fog nodes in domain $\mathcal{M}$ to their neighboring fog nodes in domain  $\mathcal{M}$. Upon reception of estimated waiting times from the Central Node, fog nodes record them in a {\em Reachability table} that they maintain. The Reachability table is utilized by fog nodes when making a decision as to which fog node to offload a request for processing. The Reachability table has three columns as it is shown in Table \ref{rt}. The information in the Reachability table of a fog node in domain $\mathcal{M}$ is updated by both the Central Node of domain $\mathcal{M}$ and the fog node itself; the Central Node announces the estimated waiting times of neighboring fog nodes, and the fog nodes measure the round-trip delay among themselves. An efficient way to estimate the waiting time of fog nodes is discussed in Section \ref{queue-status}. 
\begin{table}[!t]
		\caption{An Example of Reachability Table of Fog Node}
		\label{rt}
		\centering
		\begin{tabular}{cc|c}
			\hline
			RTT (ms) & Est. Waiting Time (ms) & Node ID\\
			\hline
			3.85 & 30.4 & 129.110.83.81 *\\
			5.3 & 72.3 & 129.110.5.75 \\
			\vdots & \vdots & \vdots\\
			\hline
		\end{tabular}
	\end{table}

Using estimated waiting time and round-trip delay in the Reachability table, each fog node selects a fog node from among its neighbors as the {\em best} fog node. It does so by selecting a neighboring fog node in the same domain with the smallest estimated waiting time plus half of the round-trip delay. The best fog node is marked with star * in Table \ref{rt}.

In the distributed mode of interaction, there is no Central Node; instead, fog nodes in domain $\mathcal{M}$ run a protocol to distribute their state information to the neighboring fog nodes in the same domain. Each fog node maintains the Reachability table using the estimated waiting time it receives from neighboring fog nodes, and the round-trip delay from itself to its neighbors. Similar to the central mode of interaction, a fog node always selects the best neighbor with the smallest estimated waiting time plus half of the round-trip delay.

{\bf Comparison}: The central mode of interaction can be seen as central resource orchestration, where the Central Node is knowledgeable of the topology and the state of the fog nodes in a domain. Inherently, the central mode of interaction is easier to implement, because there is no need for a distributed communication protocol among fog nodes; all the procedures of the estimated waiting time announcements will be implemented on the Central Node. Moreover, the Central Node could be used to push fog applications and software updates to the fog nodes in a domain.

On the other hand, the distributed mode is more suitable for scenarios in which fog nodes are not necessarily static, or when the fog network is formed in an ad hoc manner. Furthermore, in the distributed mode, there is no need to have a dedicated node to act as the Central Node, which is a reduction in the cost of deployment and less vulnerable to a single point of failure. For the purpose of this paper and our simulation studies, we have chosen the distributed mode of interaction, since our policy only needs to announce estimated waiting time updates, which is fairly easy to implement on fog nodes. 
\subsection{When to Offload a Task}  \label{when-offload}
In this subsection, we discuss the decision fog nodes make for processing or offloading a task to other fog nodes. The concept of computation offloading for mobile devices is studied in \cite{CyberForaging, offloadHeuristics}, and a number of possible policies are discussed (based on energy consumption, response time, availability, etc.). In our scheme, the decision to offload a task is based on the response time of a fog node, which depends on several factors: the amount of computation needed to be performed on a task, and the queueing status (in terms of current load) and processing capability of a fog node.  In particular, we propose a model that takes into account the different processing times of different individual tasks. In other words, in our model, there is a distinction between heavy processing tasks and light processing tasks.

\begin{figure}[!t]
	\includegraphics[width=\linewidth]{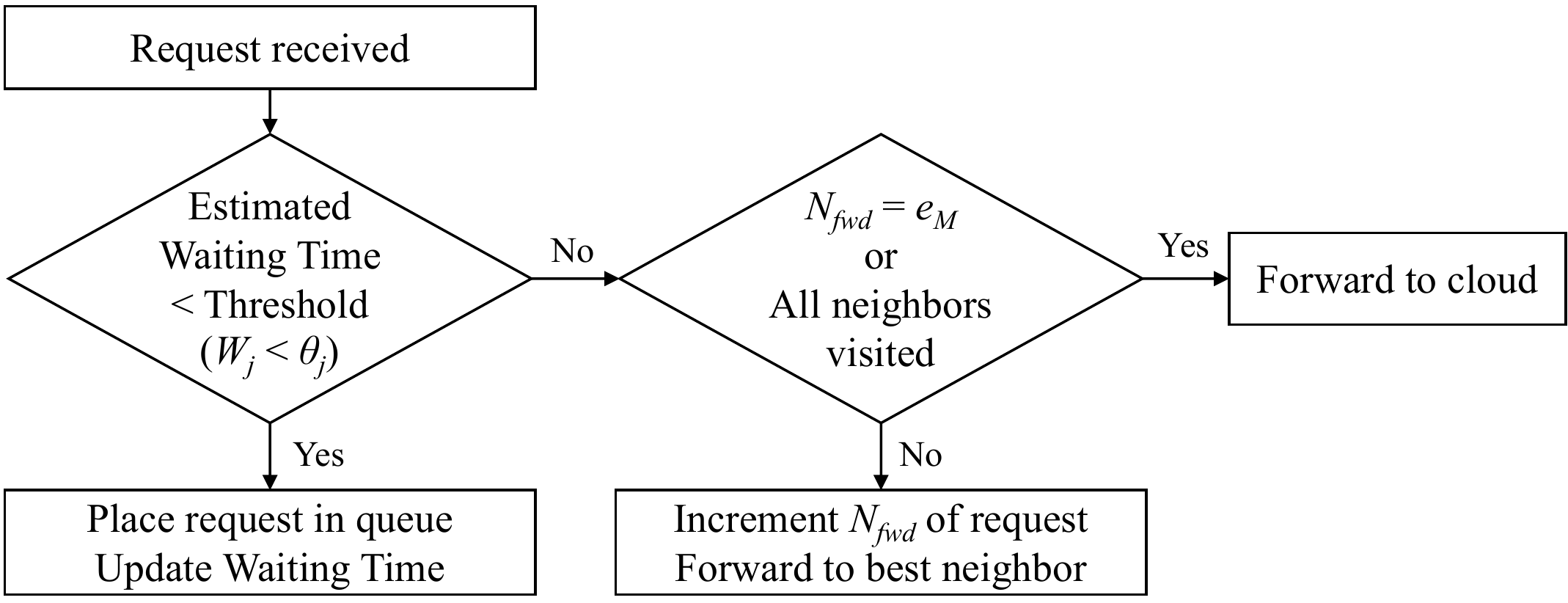}
	\caption{Policy of fog node $j$ for handling the received requests.}
	\label{fig:fog-state}
\end{figure}

We assume requests have two types: type {\em Light} (light processing) with an average processing time of $
z_j$ at the fog node $j$ and $\mathrm{Z}_k$ at the cloud server $k$, and type {\em Heavy} (heavy processing) with an average processing time of $z'_j$ at the fog node $j$ and $\mathrm{Z}'_k$ at the cloud server $k$. For example, requests sent by temperature sensors to fog nodes for calculating the average room temperature can be seen as light processing tasks. Similarly, a license plate reading request in a recorded video of a vehicle, sent by a traffic camera to fog nodes is an example of a heavy processing task. Note that, in general, more than two task types could be considered; however, in this paper, we only consider two task types for the simplicity of the presentation.

The procedure of processing or forwarding requests by fog nodes is shown in Fig. \ref{fig:fog-state}. When a fog node receives a request, it first checks the estimated waiting time of tasks in the queue.

A method for estimating processing delay and hence waiting time of the tasks in the queue is discussed in Section \ref{queue-status}. If the estimated waiting time is smaller than a threshold $\theta_j$ at fog node $j$, the fog node will accept the task. The task enters the queue and the estimated waiting time is updated. If not, the fog node offloads this request, either to one of its neighbors, or to the cloud. If the number of times the request has been offloaded ($N_{\textit{fwd}}$) is less than the {\em offload limit} $e_\mathcal{M}$ for domain $\mathcal{M}$ (the domain where fog node $j$ belongs), the request will be forwarded to a neighboring fog node. If not, it will be forwarded to the cloud.

The value of $\theta_j$ depends on the incoming traffic pattern from IoT nodes to fog nodes in a domain. In general, if all fog nodes have low load, offloading is unnecessary; and if all fog nodes have heavy load, offloading will not help to reduce the delay significantly. Offloading helps when there is a high degree of variance in the load among fog nodes. The value of $\theta_j$ could be adjusted in implementation based on the given traffic demands, to reach an optimal value that minimizes the average service delay.
	
\subsection{Finding Best Neighbor} \label{finding-best-neighbor}
Round-trip delays can be measured at the time of the setup, so that fog nodes have the values for the Reachability Table. Upon receiving an estimated waiting time sample, either from the Central Node or another fog node depending on the mode, the fog node updates the corresponding estimated waiting time in the Reachability table. As discussed, a fog node selects as its best neighbor the fog node for which the estimated delay for the current request, if offloaded, is minimal.

When a request is offloaded to the fog node's best neighbor, it undergoes roughly half of the corresponding round-trip delay to reach the best neighbor. If the request enters the best neighbor's queue, it spends time roughly equal to the estimated waiting time of the best neighbor, before it finishes processing at that neighbor. If the request does not enter the queue of the fog node's best neighbor, the request should be offloaded again (multiple offloads are possible when neighboring fog nodes are busy with requests). 

It is worth mentioning that if the fog nodes are mobile, they may need to frequently measure the round-trip delays and update them in the Reachability table. Note that for making ``optimal'' offloading decision, other components, such as propagation and transmission delay between fog nodes and IoT nodes should be considered. However, this may not be practical, since each fog node would need to know the delay from the neighboring fog nodes to the corresponding IoT nodes associated with them. 
\subsection{Checking and Updating Queue Status}  \label{queue-status}
When fog nodes receive requests from IoT nodes that participate in an application, they need to distinguish between type Light and type Heavy requests, in order to update the queue parameters. To address this, we assume requests have in their header a field that identifies the type of the request (for instance {\em Traffic Class} field in IPv6 header). The application hence sets the field in the header of the packets being generated from IoT nodes.

Fog nodes always need to know an estimate of the current total processing delay of the tasks in their queue (i.e. estimated waiting time), both for when they need to make offloading decisions, and when reporting their estimated waiting time to neighboring fog nodes. A possible method for estimating the waiting time is for the fog node to store an estimate of the {\em current} waiting time of the tasks in the queue. Upon arrival to or departure from the queue, the fog node simply updates the estimated waiting time of the tasks in the queue.
\begin{table}[!b]
	\newcolumntype{L}[1]{>{\raggedright\let\newline\\\arraybackslash\hspace{0pt}}m{#1}}
	\caption{Table of Notations}
	\centering
	{\footnotesize
		{\renewcommand\arraystretch{1.0} %scale the height of table
			\begin{tabular}{c|L{0.75\columnwidth}}
				\hline
				$d_i$ & Service delay for an IoT node $i$\\
				%$d_i^{\max}$ & Maximum delay allowed by application for IoT node $i$\\
				\hline
				$p^I_i$ & Probability that IoT node $i$ processes its own request\\
				$p^F_i$ & Probability that IoT node $i$ sends its request to fog layer\\
				$p^C_i$ & Probability that IoT node $i$ sends its request to the cloud\\
				\hline
				$X_{st}^{LL'}$ & Propagation delay from node $s$ in layer $L$ to node $t$ in layer $L'$, where $s, t \in \{ i, j, k \}$ and $L, L' \in \{ I, F, C \}$\\
				
				$Y_{st}^{LL'}$ & Sum of all transmission delays on links between node $s$ in layer $L$ to node $t$ in layer $L'$, where $s, t \in \{ i, j, k \}$ and $L, L' \in \{ I, F, C \}$\\
				\hline
				$A_i$ & Average processing delay of requests at IoT node $i$\\
				$a_i$ & Average processing delay of type Light requests at IoT node $i$ ($a'_i$ is average proc. delay of type Heavy requests)\\
				\hline
				$L_{ij}$ & Delay of processing and handling requests of IoT node $i$ in the fog layer (and possibly the cloud layer), where fog node $j$ is the fog node to which IoT node $i$ initially sends its request ($L_{ij}=L_{ij}(0)$)\\
				$L_{ij}(x)$ & Delay of processing and handling requests of IoT node $i$ in fog layer (and possibly cloud layer), by fog node $j$ during the $x$'th offload in the fog layer\\
				\hline
				$S^L_D$ & Set of nodes in domain $D$ at layer $L$, where $(L, D) \in \{ (I, \mathcal{P}), (F, \mathcal{M}), (C, \mathcal{N}) \}$\\
				$S^L$ & $\bigcup_{D} S^L_D$ : set of nodes (in all domains) at layer $L$\\
				\hline
				$\overline{H}_k$ & Average waiting time at cloud server $k$\\
				$\overline{\Delta}_k$ & Average waiting time of a single processing unit at cloud server $k$\\
				\hline
				$\varsigma_i$ & Average size of request data that IoT node $i$ generates\\
				\hline
				$b_i$ & Probability that a generated request at IoT node $i$ is Light\\
				\hline
				$W_j$ & Waiting time of fog node $j$\\
				$c_j$ & Number of type Light requests in fog node $j$'s queue\\ 
				\hline
				$P_j$ & Probability that an incoming request is accepted by fog node $j$\\
				\hline
				$\theta_j$ & Offloading threshold at fog node $j$\\
				\hline
				$e_\mathcal{M}$ & Maximum offload limit at the fog layer in domain $\mathcal{M}$\\
				\hline
				$q$ & The fog fairness parameter\\
				\hline
			\end{tabular}
		}}
		\label{parameters}
	\end{table}
To calculate the estimated waiting time of the tasks in the queue $W$, fog node $j$ periodically measures processing times of recently processed tasks ($new\_z_j$) and updates the estimated processing time of each task type ($z_j$). For light processing tasks, we have $z_j=(1-\alpha)\cdot z_j + \alpha\cdot new\_z_j$ (a similar equation holds for heavy processing tasks). This equation is a weighted average that maintains a weight of $\alpha$ for new measured processing times and a weight of $1-\alpha$ for the old measurements (current estimate).

To obtain the total processing time of the tasks in the queue on the fly, fog node $j$ stores the current number of type Light and Heavy requests in the queue $c_j$ and $c'_j$, respectively, in addition to $z_j$ and $z'_j$. Fog nodes then multiply the estimated processing time of each task type by the number of tasks of that type in the queue. In other words, the estimated waiting time $W$ of the fog node $j$ will be
\begin{IEEEeqnarray}{rCl}  \label{waiting-time}
	W_j = c_j\cdot z_j + c'_j\cdot z'_j.
\end{IEEEeqnarray}
\section{Analytical Model} \label{analytical}
\subsection{Network Model}
We model the network as an undirected graph $G=[V;E;w]$, where node set $V$ includes set of IoT nodes, fog nodes, and cloud servers ($V=S^I\cup S^F\cup S^C$). The Edge set $E$ represents the communication links between the nodes. For instance, there is a link between IoT node $i$ and fog node $j$ if they communicate. The edge weight set $w$ represents the weight of the edges between nodes. We use the tuple $(X,R)$ for the edge weights, where $X$ represents the propagation delay, and $R$ the transmission rate between two nodes. We will not pose any restrictions on the topology, except for the mentioned logical three-layer architecture, as this makes the model easy to present.
\subsection{Service Delay}
Recall that IoT nodes process requests locally, send it to a fog node, or send it to the cloud. Thus, service delay $d_i$ for an IoT node $i$ can be written as:
 \begin{IEEEeqnarray}{rCl} \label{service_delay}
 	d_i&=&p^I_i\cdot (A_i) + p^F_i\cdot (X_{ij}^{IF}+Y_{ij}^{IF} + L_{ij})\IEEEnonumber\\
 	&&+ p^C_i\cdot (X_{ik}^{IC}+Y_{ik}^{IC}+\overline{H}_{k}+X^{CI}_{ki}+Y^{CI}_{ki});\IEEEnonumber\\
 	&&~~ j=f(i),~k=g(i),
 \end{IEEEeqnarray}
where $p^I_i$ is the probability that the IoT node $i$ processes its own request at the IoT layer, $p^F_i$ is the probability of sending the request to the fog layer, and $p^C_i$ is the probability that the IoT node sends the request directly to the cloud; $p^I_i + p^F_i + p^C_i = 1$. $A_i$ is the average processing delay of the IoT node $i$ when it processes its own request. $X_{ij}^{IF}$ is propagation delay from IoT node $i$ to fog node $j$, $Y_{ij}^{IF}$ is sum of all transmission delays on links from IoT node $i$ to fog node $j$. Similarly, $X_{ik}^{IC}$ is propagation delay from IoT node $i$ to cloud server $k$, $Y_{ik}^{IC}$ is sum of all transmission delays on links from IoT node $i$ to cloud server $k$. $X^{CI}_{ki}$ and $Y^{CI}_{ki}$ are the propagation and transmission delays from cloud server $k$ to IoT node $i$.
 
The transmission and propagation delay from the fog layer to IoT node $i$ will be included in $L_{ij}$, since the request may be further offloaded to a different node in the fog layer (more details in Section \ref{delay_fog}).

$L_{ij}$ is the delay for processing and handling requests of IoT node $i$ in the fog layer (and possibly cloud layer, if fog nodes offload the request to the cloud), where fog node $j$ is the first fog node to which IoT node $i$ initially sends its request. Note that fog node $j$ might offload the request to another fog node or to the cloud, and that all the corresponding incurred delays are included in $L_{ij}$. $\overline{H}_k$ is the average delay for handling the request at the cloud server $k$, which consists of the queueing time at the cloud server $k$ plus the processing time at the cloud server $k$. ($L_{ij}$ and $\overline{H}_k$ will be discussed in further detail in Sections \ref{delay_fog} and \ref{delay_cloud} respectively).
 
$f(i)$ and $g(i)$ are mapping functions that indicate the fog node $j$ and cloud server $k$ to which IoT node $i$ sends its requests, respectively (refer to Table \ref{table_rates}). For instance, if in an application, IoT node $i$ always sends its requests to fog node $j^*$ in the fog layer, then $f(i)=j^*$. In another scenario if IoT nodes always send their requests to the closest fog node in the fog layer, then $f(i)=\arg\min_j X_{ij}^{IF}$, which translates to the index of the fog node with smallest propagation delay (distance) from IoT node $i$. 

To formalize the problem further, let us define an IoT-fog-cloud application $\Psi$. In the rest of this work all the equations are defined on a single application $\Psi(\mathcal{N},\mathcal{M},\mathcal{P})$. 
 \begin{definition}
 	IoT-fog-cloud {\em application} $\Psi$ is an application on domain $\mathcal{N}$ of cloud servers, domain $\mathcal{M}$ of fog nodes and domain $\mathcal{P}$ of IoT nodes, and is written as $\Psi(\mathcal{N},\mathcal{M},\mathcal{P})$. Examples of $\Psi$ are: video processing, temperature sensor reporting, traffic road analysis, and oil rig pressure monitoring.
 \end{definition}
 We do not assume any distribution for $p^I_i$, $p^F_i$, and $p^C_i$, since their values will be defined by individual applications and based on QoS requirements and policies. In other words, they will be given to the scheme as input. In Section \ref{results}, we assume different values for these probabilities for type Heavy and Light requests, depending on the type of IoT node $i$. 
 %For example, in an application $\Psi(\mathcal{N},\mathcal{M},\mathcal{P})$, if the chances that IoT node $i\in \mathcal{P}$ sends requests to fog layer is 74\% and to the cloud layer is 15\%, then $p^I_i=0.11$, $p^F_i=0.74$, and $p^C_i=0.15$. 
 
By using the model to evaluate the service delay for the IoT nodes in one application domain, our goal is to minimize delay through the definition of policies for exchanging requests between IoT nodes, fog nodes, and cloud servers. We formulate the above statement as the minimization of the average service delay of IoT nodes in domain $\mathcal{P}$, or
 \begin{IEEEeqnarray}{rCl} 
 	&\min& \frac{1}{|S_\mathcal{P}^{I}|}\sum_{i\in S_\mathcal{P}^{I}}{d_i},
 \end{IEEEeqnarray} 
where $S_\mathcal{P}^{I}$ denotes the set of IoT nodes that are in domain $\mathcal{P}$. Similar to above, $S_\mathcal{M}^{F}$ indicates the set of fog nodes in domain $\mathcal{M}$, and $S_\mathcal{N}^{C}$ is the set of cloud servers in domain $\mathcal{N}$. In the following subsection, we will discuss in more details the components of the service delay. 
\subsection{Propagation and Transmission Delays} 
In Equation (\ref{service_delay}) we have the IoT-cloud delay terms $X_{ik}^{IC}$, $Y_{ik}^{IC}$, $X_{ki}^{CI}$, $Y_{ki}^{CI}$ when the IoT node sends its requests directly to the cloud layer. These terms are effective in the equation in cases where the application $\Psi$ is not implemented in the fog layer ($S_\mathcal{M}^{F}=\{\}$), or when the request must be sent to the cloud for archival purposes, or when there is no fog layer and the IoT node communicates directly to the cloud.

Recall that $Y_{ij}^{IF}$ is the sum of all transmission delays on the links from IoT node $i$ to fog node $j$. If IoT node $i$ and fog node $j$ are $l$-hop neighbors, we will have 
\begin{equation}
Y_{ij}^{IF}=\sum_l{\frac{\varsigma_i}{R_l}} .
\end{equation}
where $\varsigma_i$ is the average amount of request data that IoT node $i$ generates, and $R_l$ is the transmission rate of the $l$'th link between IoT node $i$ and fog node $j$. The expanded equations for transmission delays between other layers ($Y_{ik}^{IC}$, $Y_{jk}^{FC}$, etc.) are derived similarly to $Y_{ij}^{IF}$.
\subsection{Processing Delay of IoT node} 
As explained in Equation (\ref{service_delay}), for IoT node $i$, the average processing delay is $A_i$. If $b_i$ denotes the probability that a generated request at IoT node $i$ is type Light, and $b'_i=1-b_i$ is the probability that a  generated request at IoT node $i$ is type Heavy, $A_i$ could be written as 
\begin{equation}
	A_i=b_i\cdot a_i+b'_i\cdot a'_i ,
\end{equation}
where $a_i$ is the average processing time of requests of type Light at IoT node $i$, and $a'_i$ is the average processing time of requests of type Heavy at IoT node $i$. If IoT node $i$ is of type Light (or type Heavy), i.e. it only generates type Light (or type Heavy) requests, $b_i=1$ (or $b_i=0$) and $A_i=a_i$ (or $A_i=a'_i$). Should more than two types of tasks be considered, the equation above (and other equations with two task types) could be generalized to support more task types. 

Table \ref{table_rates} summarizes the important parameters of different layers and shows the mapping functions that are used in the analytical model. A few of the parameters and mapping functions have already been introduced (such as $a_i$ and $f(i)$); yet others will be introduced in the following subsections. 
\subsection{Delay in Fog Layer} \label{delay_fog}
In this section, we define a recursive equation for $L_{ij}$. Let us define $L_{ij}(x)$ as the delay of processing and handling requests of IoT node $i$ in the fog layer (and possibly the cloud layer), by fog node $j$ during the $x$'th offload in the fog layer ($x \ge 0$). Also, let us label $L_{ij} \equiv L_{ij}(0)$. If $P_{j}$ denotes the probability that a request is accepted by fog node $j$ (enters the queue of fog node $j$)  $L_{ij}(x)$ can be written as:
\begin{IEEEeqnarray}{rCl} \label{LM}
	L_{ij}(x)&&=P_{j}\cdot(\overline{W}_j+X^{FI}_{ji}+Y^{FI}_{ji})\IEEEnonumber\\
	&&+(1-P_{j})\cdot\bigg{[}[1-\phi(x)].\big{[}X^{FF}_{jj'}+Y^{FF}_{jj'}+L_{ij'}(x+1)\big{]}\IEEEnonumber\\
	&&~~~~~~~+\phi(x)\cdot\big{[}X^{FC}_{jk}+Y^{FC}_{jk}+\overline{H}_k+X^{CI}_{ki}+Y^{CI}_{ki}\big{]}\bigg{]};\IEEEnonumber\\
	&&~~ j'=\textit{best}(j),~k=h(j).
\end{IEEEeqnarray}
In the equation above, $\overline{W}_j$ is the average waiting time in fog node $j$. $\phi(x)$ is the offloading function, which is defined as 
\begin{equation} 
\phi(x) =
\begin{cases}
0 & x<e_\mathcal{M}\\
1 & x=e_\mathcal{M}\\
\end{cases}.
\end{equation}
If $x<e_\mathcal{M}$, then $\phi(x)=0$, which indicates that the request will be offloaded to another fog node. If $x=e_\mathcal{M}$, then $\phi(x)=1$, which means that the forward limit is reached and that the request will be offloaded to the cloud (recall Fig. \ref{fig:fog-state}). $x$ takes on integer values in $[0, e_\mathcal{M}]$.

$\textit{best}(j)$ and $h(j)$ are mapping functions that map a particular fog node $j$ to its best fog neighbor and the cloud server associated with the fog node, respectively (see Table \ref{table_rates}). Since the choice of the best neighbor of a fog node depends on the current state of the system, the system is dynamic and $best(j)$ will be a pointer to the current best neighbor of fog node $j$. $h(j)$ simply holds the index of the associated cloud server for fog node $j$.

{\bf Explanation}: Assume fog node $j$ is the one that is selected first by the IoT node $i$. When a request from an IoT $i$ node reaches the fog layer, fog node $j$ first tries to process the request. The request enters this node's processing queue with probability $P_{j}$, and does not with probability $(1-P_{j})$, which depends on estimated waiting time. If the request enters the queue, it will experience average waiting time $\overline{W}_j$, and propagation and transmission delays of $X^{FI}_{ji}$ and $Y^{FI}_{ji}$ to return back to the IoT node. Note that the processing delay of the current task entering the fog node $j$'s queue is already included in the calculation of $\overline{W}_j$, because of the way waiting time is defined.

If the request does not enter fog node $j$, fog node $j$ will offload the request to its best fog neighbor $j'$, which incurs a propagation and transmission delay of $X^{FF}_{jj'}$ and $Y^{FF}_{jj'}$, respectively. The request also undergoes a delay of $L_{ij'}(x+1)$, which is the delay of processing and handling the request in the fog layer (and possibly the cloud layer), by fog node $j'$ during the $x+1$'st offload. Finally when a request has been offloaded $e_\mathcal{M}$ times ($\phi(x)=1$), if the last fog node needs to offload the request, it will do so by offloading it to the cloud, which incurs fog-cloud propagation and transmission delay of $X^{FC}_{jk}$ and $Y^{FC}_{jk}$, respectively, cloud processing delay of $\overline{H}_k$, and cloud-IoT propagation and transmission delay of $X^{CI}_{ki}$ and $Y^{CI}_{ki}$, respectively.

The reason $X^{FI}_{ji}$ and $Y^{FI}_{ji}$ are included in Equation (\ref{LM}) and not Equation (\ref{service_delay}) is because a request sent from IoT node $i$ to fog node $j$ could be received from fog node $j'$ (offloaded to and processed at fog node $j'$). In this case the propagation and transmission delay from IoT layer to fog layer are $X^{IF}_{ij}$ and $Y^{IF}_{ij}$, respectively, but the propagation and transmission delays from fog layer to IoT layer are $X^{FI}_{j'i}$ and $Y^{FI}_{j'i}$.

{\bf Boundary case}: Consider a domain $\mathcal{M}$ of fog nodes where $e_\mathcal{M}=0$, which means no forwarding is allowed. In this case, if a request does not enter a fog node's queue, it will be offloaded to the cloud. In this case, $L_{ij} = P_{j}\cdot(\overline{W}_j+X^{FI}_{ji}+Y^{FI}_{ji}) + (1-P_{j})\cdot[X^{FC}_{jk}+Y^{FC}_{jk}+\overline{H}_k+X^{CI}_{ki}+Y^{CI}_{ki}]$.
\begin{table}[!t]
	\renewcommand\arraystretch{1} 
	\caption{Parameters and Mapping Function of Different Layers}
	\centering
	\resizebox{0.5\textwidth}{!}{
		\begin{tabular}{|c||c|c|c|c|c|}
			\hline
			
			\textbf{Layer}&\textbf{Mapping}&\textbf{Node}&\textbf{Arrival}&\textbf{Service}&\textbf{Avgerage}\\
			\textbf{Name}&\textbf{Function}&\textbf{Index}&\textbf{Rate}&\textbf{Rate}&\textbf{Processing Time}\\[1ex]
			\hline
			\hline
			Cloud&\multirow{3}{40px}{\includegraphics[scale=0.75]{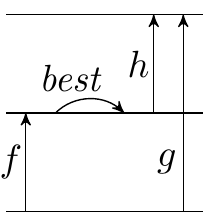}}&$k$&$l_k$, $l'_k$&$u_k$, $u'_k$&$\mathrm{Z}_k/m_k$, $\mathrm{Z}'_k/m_k$\\[3ex]
			
			Fog&&$j$&$\lambda_j$, $\lambda'_j$&$\mu_j$, $\mu'_j$&$z_j$, $z'_j$\\[3ex]
			
			IoT&&$i$&$\gamma^{}_ip^I_i$, $\gamma'_ip^I_i$&$\nu_i$, $\nu'_i$&$a_i$, $a'_i$\\[1ex]
			\hline
			
		\end{tabular}
	}
	\label{table_rates}
\end{table}

% {\bf Special case}: Consider a scenario where a fog node sends some sort of feedback message to the cloud, when processing tasks from IoT nodes (for example aggregated temperature reading for archival purposes). Note that this does not add any new term in the delay equations from the perspective of IoT nodes, because sending feedback messages to the cloud does not affect the service delay experienced by an IoT node. 
\subsection{Average Waiting Time of Fog Node}
To obtain an equation for the average waiting time of fog node $j$, $\overline{W}_j$, we need to model fog nodes. We assume a fog node is a single server with a large enough queue to hold infinite incoming requests. The incoming traffic to fog nodes is considered to be Poisson (request type Light with rate $\lambda_{j}$ and request type Heavy with rate $\lambda'_{j}$) and processing times to be exponentially distributed (request type Light with rate $\mu_{j}$ and request type Heavy with rate $\mu'_{j}$). Thus, fog nodes can be modeled as Markovian queuing systems with multi-class traffic. 

To derive the equation for average waiting time of fog node, we need to define the state of the system. We define $P^j_{n,n'}$ as the state in which there are $N=n$ type Light requests and $N'=n'$ type Heavy requests in the fog node $j$ (i.e. $n+n'-1$ requests are in the queue and 1 request is in service, if $n,n'\neq0$). Thus $\overline{W}_j$ can be calculated as: 
\begin{equation}
	\overline{W}_j=E(W_j)=\sum_{n}{\sum_{n'}{E(W_j|N=n, N'=n') P^j_{n,n'}}} ,
\end{equation}
where random variable $W_j$ denotes the waiting time of fog node $j$. We will derive a closed form equation for $P^j_{n,n'}$ in Section \ref{steady-state-prob}. $E(W_j|N=n, N'=n')$ is calculated as: 
\begin{equation}
	E(W_j|N=n, N'=n')=n\cdot\frac{1}{\mu_j}+n'\cdot\frac{1}{\mu'_j},
\end{equation}
and hence $\overline{W}_j=\sum_{n}{\sum_{n'}{[(\frac{n}{\mu_j}+\frac{n'}{\mu'_j})P^j_{n,n'}}]}$.
\subsection{Fog Node Steady State Probability $P^j_{n,n'}$} \label{steady-state-prob}
To obtain the fog steady state probability $P^j_{n,n'}$, one can model the system as a two-dimensional Markov chain and solve for steady-state probabilities $P^j_{n,n'}$. The states are labeled as $(n,n')$, denoting $n$ type Light and $n'$ type Heavy requests in the queue. If $r_{t:t'}$ signifies the transition rate from state $t$ to $t'$, we can obtain the transition rates of the state diagram as follows:
\begin{IEEEeqnarray}{rCl}
	r_{(n,n'):(n+1,n')} &=& \lambda_j ,\\ 
	r_{(n,n'):(n,n'+1)} &=& \lambda'_j ,\\
	r_{(n,n'):(n-1,n')} &=& Q_{n,n'} \mu_j ,\\
	r_{(n,n'):(n,n'-1)} &=& (1-Q_{n,n'}) \mu'_j ,
\end{IEEEeqnarray}
where $Q_{n,n'}$ is a state-dependent function that is defined as
\begin{equation}
Q_{n,n'}=\frac{q n}{q n+(1-q) n'}.
\end{equation}
In the above definition, $q\in (0,1)$ is the fairness parameter and its value depends on how the fog node selects jobs from its queue. 

Fog nodes segregate type Light and Heavy requests in their queue, and depending on the value of $q$, select the next job to process from each of the two groups. The closer the value of $q$ is to $0$ (or $1$), the higher priority is given to type Heavy (or type Light) requests in the queue for processing. When $q=0.5$, $Q_{n,n'}=\frac{n}{n+n'}$, and this is equivalent to the case when the fog node simply selects the head of the queue for processing. More specifically, for $q=0.5$, if the fog node is in state $(n,n')$, on average the system processes a type Light request and goes to state $(n-1,n')$ with rate $\frac{n}{n+n'}\mu_j$, and processes a type Heavy request and goes to state $(n,n'-1)$ with rate $\frac{n'}{n+n'}\mu'_j$. This is the case because on average the head of the queue is a Light request with probability $\frac{n}{n+n'}$ and is a Heavy request with probability $\frac{n'}{n+n'}$. 
\subsection{Acceptance Probability: $P_j$}
$P_j$ is the probability that a request is accepted by fog node $j$ (enters the queue of fog node $j$) and is used in Equation (\ref{LM}). $P_j$ depends on the queuing state of a fog node; in particular, if fog node $j$'s estimated waiting time is greater than a threshold $\theta_j$, it will offload the request to its best neighbor. Thus $P_j$ is extended by the following probability:
\begin{IEEEeqnarray}{rCl} \label{p-j}
	P_j&=&P[\text{request enters the queue at fog node }j]\\
	&=&P[\text{waiting time of fog node }j<\theta_j]=P[W_j<\theta_j].\IEEEnonumber
\end{IEEEeqnarray}

To evaluate the equation above, we need to derive the probability density function (PDF) of waiting time $W_j$. Recall that waiting time $W_j$ is the sum of the processing delays of all the requests in fog node $j$. Let the random variables $x^j_l$ and $y^j_l$ denote the processing delays of the $l$'th request of type Light and type Heavy in fog node $j$, respectively. If there are $N$ type Light and $N'$ type Heavy requests in the fog node $j$, the waiting time is  $W_j=\sum_{l=1}^{N}{x^j_l}+\sum_{l=1}^{N'}{y^j_l}$. 

Let $X^j_{n}=\sum_{l=1}^{n}{x^j_l}$, which is the sum of $n$ processing delays (exponentially distributed) of request type Light, and $Y^j_{n'}=\sum_{l=1}^{n'}{y^j_l}$, which is the sum of $n'$ processing delays (exponentially distributed) of request type Heavy. Thus $W_j=X^j_N+Y^j_{N'}$. Note that the sum of $m$ independent and identically distributed exponential random variables with parameter $\mu$ is a gamma random variable with parameters $(m,\mu)$. Hence, the PDF of $X^j_{n}$ and $Y^j_{n'}$ will follow gamma distributions as follows:
\begin{IEEEeqnarray}{rCl}
	f_{X^j_n}(t)=\frac{\mu_j(\mu_jt)^{n-1}.e^{-\mu_jt}}{(n-1)!}u(t),\\ f_{Y^j_{n'}}(t)=\frac{\mu'_j(\mu'_jt)^{n'-1}.e^{-\mu'_jt}}{(n'-1)!}u(t),
\end{IEEEeqnarray}
such that $u(t)$ is the unit step function. Note that the shown PDFs are gamma distributions with parameters $(n,\mu_j)$ and $(n',\mu'_j)$, respectively. What follows is the derivation of $P_j$:
\begin{IEEEeqnarray}{rCl} \label{p-j-expanded}
	P_j&=&P[W_j<\theta_j]=P[X^j_N+Y^j_{N'}<\theta_j]\IEEEnonumber\\
	&=&\sum_{n=0}^{\infty}{\sum_{n'=0}^{\infty}{P[X^j_N+Y^j_{N'}<\theta_j|N=n,N'=n']P^j_{n,n'}}}\IEEEnonumber\\
	&=&P^j_{0,0}+\sum_{n=1}^{\infty}{P[X^j_n<\theta_j]P^j_{n,0}}+\sum_{n'=1}^{\infty}{P[Y^j_{n'}<\theta_j]P^j_{0,n'}}\IEEEnonumber\\
	&+&\sum_{n=1}^{\infty}{\sum_{n'=1}^{\infty}{P[X^j_n+Y^j_{n'}<\theta_j]P^j_{n,n'}}}\IEEEnonumber\\
	&=&P^j_{0,0}+\sum_{n=1}^{\infty}{[\int_{0}^{\theta_j}{\negspace f_{X^j_n}(t)dt}]P^j_{n,0}}+\sum_{n'=1}^{\infty}{[\int_{0}^{\theta_j}{\negspace f_{Y^j_{n'}}(t)dt}]P^j_{0,n'}}\IEEEnonumber\\
	&+&\sum_{n=1}^{\infty}{\sum_{n'=1}^{\infty}}{[\int_{0}^{\theta_j}{\negspace f_{X^j_n+Y^j_{n'}}(t)dt}]P^j_{n,n'}}
\end{IEEEeqnarray}
In order to expand the summation on the second line, we separated the sum into four cases in the following order: ($n=0$, $n'=0$), ($n>0$, $n'=0$), ($n=0$, $n'>0$), and ($n>0$, $n'>0$). 

We have the equations for $f_{X^j_n}(t)$ and $f_{Y^j_{n'}}(t)$, but we do not have an equation for $f_{X^j_n+Y^j_{n'}}$. Since $X^j_{n}$ and $Y^j_{n'}$ are independent, the PDF of $X^j_n+Y^j_{n'}$ will be the convolution of $f_{X^j_n}(t)$ and $f_{Y^j_{n'}}(t)$. We transform $f_{X^j_n}(t)$ and $f_{Y^j_{n'}}(t)$ to their corresponding Laplace transforms $\mathcal{L}\{f_{X^j_n}\}$ and $\mathcal{L}\{f_{Y^j_{n'}}\}$, so that the Laplace transform of $X^j_n+Y^j_{n'}$ is the product of $\mathcal{L}\{f_{X^j_n}\}$ and $\mathcal{L}\{f_{Y^j_{n'}}\}$. We have
\begin{equation}
	\mathcal{L}\{f_{X^j_n}\}(s) = \frac{(\mu_j)^n}{(s+\mu_j)^n}, ~~~  \mathcal{L}\{f_{Y^j_{n'}}\}(s) = \frac{(\mu'_j)^{n'}}{(s+\mu'_j)^{n'}}.
\end{equation}
The PDF of $X^j_n+Y^j_{n'}$ is then given by:
\begin{equation}
	\negspace f_{X^j_n+Y^j_{n'}}(t)=(\mu_j)^n(\mu'_j)^{n'}\mathcal{L}^{-1}\{\frac{1}{(s+\mu_j)^n(s+\mu'_j)^{n'}}\}.
\end{equation}
We now have all the components of Equation (\ref{p-j-expanded}). In the following section, we discuss how to obtain the arrival rates to the fog nodes.
\subsection{Arrival Rates to Fog Nodes}
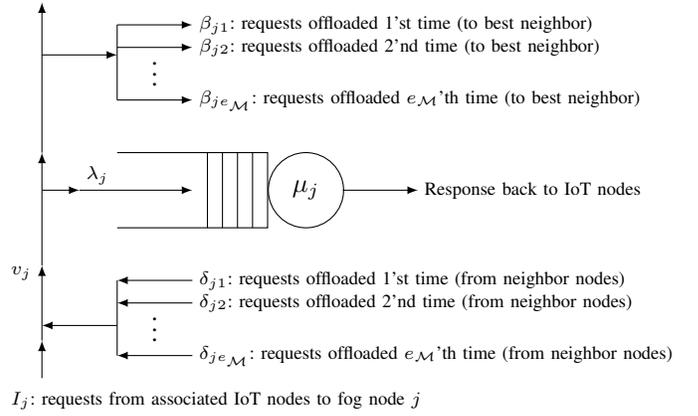
\begin{figure}[!t]  
	\begin{tikzpicture} 
	\def\server{(2.51,0.5)}
	
	\tikzset{
		%Define standard arrow tip
		>=stealth',
		% Define arrow style
		pil/.style={
			->,
			thick,
			shorten <=2pt,
			shorten >=2pt,}
	}
	
	% The server and the queue
	\draw [fill=white] \server circle [radius=0.5];
	\draw node at \server {$\mu_j$};
	\draw (0,0) -- (2,0) -- (2,1) -- (0,1);
	\draw (1.8,1) -- (1.8,0);
	\draw (1.6,1) -- (1.6,0);
	\draw (1.4,1) -- (1.4,0);
	\draw (1.2,1) -- (1.2,0);

	% the traffic
	\begin{scope}[>=latex]
	\draw [->] (3,0.5) -- (4,0.5);
	
	\draw [->] (-1,0.5) -- (-0.5,0.5);
	\draw [->] (-0.5,0.5) -- (1,0.5);
	
	%arrows up
	\draw (-1,-0.5) -- (-1,0.5);
	\draw [->] (-1,-1.5) -- (-1,-0.5);
	\draw [->] (-1,-2) -- (-1,-1.5);
	\draw [->] (0,-1.3) -- (-1,-1.3);
	
	\draw (0,-0.7) -- (0,-1.7);
	\draw [->] (1,-0.7) -- (0,-0.7);
	\draw [->] (1,-1) -- (0,-1);
	\draw [->] (1,-1.7) -- (0,-1.7);
	\draw node at (0.5,-1.25) {\vdots};
	
	\draw [->] (-1,0.5) -- (-1,1);
	\draw [->] (-1,1) -- (-1,3);
	
	\draw [->] (-1,2.3) -- (0,2.3);
	
	\draw (0,1.7) -- (0,2.7);
	\draw [->] (0,1.7) -- (1,1.7);
	\draw [->] (0,2.4) -- (1,2.4);
	\draw [->] (0,2.7) -- (1,2.7);
	\draw node at (0.5,2.15) {\vdots};
	\end{scope}
	
	{\scriptsize
		%texts
		\draw node[right] at (4,0.5) {Response back to IoT nodes};
		
		\draw node[right] at (1,-0.7) {$\delta_{j1}$: requests offloaded 1'st time (from neighbor nodes)};
		\draw node[right] at (1,-1.0) {$\delta_{j2}$: requests offloaded 2'nd time (from neighbor nodes)};
		\draw node[right] at (1,-1.7) {$\delta_{je_\mathcal{M}}$: requests offloaded $e_\mathcal{M}$'th time (from neighbor nodes)};
		
		\draw node[right] at (-1.5,-2.3) {$I_j$: requests from associated IoT nodes to fog node $j$};
		
		\draw node[right] at (1,2.7) {$\beta_{j1}$: requests offloaded 1'st time (to best neighbor)};
		\draw node[right] at (1,2.4) {$\beta_{j2}$: requests offloaded 2'nd time (to best neighbor)};
		\draw node[right] at (1,1.7) {$\beta_{je_\mathcal{M}}$: requests offloaded $e_\mathcal{M}$'th time (to best neighbor)};
		
		\draw node[right] at (-1.5,3.3) {$C_j$: requests offloaded to cloud from fog node $j$};
		
		%nodes
		%\draw node[right] at (-0.8,0.2) {$P_j$};
		%\draw node[right] at (-2,0.8) {$1-P_j$};
		\draw node[right] at (-1.5,-0.6) {$v_j$};
		\draw node[right] at (-0.5,0.7) {$\lambda_j$};
	}
	\end{tikzpicture}
	
	\caption{The traffic model of fog node $j$ (shown only for type Light requests) \label{queueing-model-fog}}
\end{figure}

The rates of arrival to fog nodes ($\lambda_j$ and $\lambda'_j$) are needed for the evaluation of the fog steady-state probability $P^j_{n,n'}$. We can obtain these rates by considering the queueing model of the fog nodes. Figure \ref{queueing-model-fog} shows the queueing model of fog node $j$ for type Light requests. Note that in this subsection we only perform analysis for obtaining equations for type Light requests, including $\lambda_j$. The equation
for $\lambda'_j$ is derived similarly to that of $\lambda_j$.

The incoming type Light traffic from associated IoT nodes to fog node $j$ is denoted by $I_j$. The offloaded type Light requests from neighboring fog nodes to fog node $j$ are labeled $\delta_{j1}$, $\delta_{j2}$, \ldots, $\delta_{je_\mathcal{M}}$. These all account for the incoming type Light traffic to fog node $j$. The incoming type Light traffic to fog node $j$ is labeled as $v_j$. Thus,
\begin{equation}
v_j = I_j+\sum_{l=1}^{e_\mathcal{M}}{\delta_{jl}}.
\end{equation}

Recall that $P_j$ is the probability that an incoming request is accepted by fog node $j$. So $\lambda_j$, the arrival rate (type Light) of fog node $j$ could be obtained by the following equation:
\begin{equation} \label{arrival-rate}
\lambda_j=P_j\cdot v_j=P_j(I_j+\sum_{l=1}^{e_\mathcal{M}}{\delta_{jl}}).
\end{equation}
When the request is not accepted by fog node $j$, the fog node $j$ offloads it to its best neighbor, or to the cloud. The offloaded type Light requests from fog node $j$ to its best neighbor are labeled $\beta_{j1}$, $\beta_{j2}$, \ldots, $\beta_{je_\mathcal{M}}$. The offloaded type Light requests from fog node $j$ to the cloud is labeled $C_j$.

In order to evaluate the Equation (\ref{arrival-rate}), we need to have closed-form equations for $I_j$ and $\delta_{jl}$'s. $I_j$ is obtained easily as follows. If $f^{-1}(j)$ denotes the mapping function that indicates the set of IoT nodes that send their requests to fog node $j$, then $I_j$ is:
\begin{equation}\label{incoming-traffic-from-iot}
I_j = \negspace \sum_{i\in f^{-1}(j)}{\negspace[\gamma^{}_i\cdot p_i^F]},
\end{equation}
where $\gamma^{}_i$ denotes the rate of generating Light requests from IoT node $i$ (in units of Erlangs). We assume IoT node $i$ generates type Light (or Heavy) requests according to a Poisson process, with rate $\gamma^{}_i$ (or $\gamma'_i$), depending on the type of IoT node $i$. 
In order to obtain equations for $\delta_{jl}$'s, we need to solve the following system of equations:
\begin{IEEEeqnarray}{rCl}
\beta_{j1}&=&(1-P_j)\cdot I_j~,~\label{offload-from-iot}\\
\beta_{j2}&=&(1-P_j)\cdot \delta_{j1}~,~\\
&\vdots&\IEEEnonumber\\
\beta_{je_\mathcal{M}}&=&(1-P_j)\cdot \delta_{j(e_\mathcal{M}-1)}.
\end{IEEEeqnarray}

If fog node $j$ cannot accept a request sent from IoT nodes ($I_j$), the request will be offloaded for the first time to fog node $j$'s best neighbor (Equation (\ref{offload-from-iot})). Similar to this explanation, other equations could be realized: if an $l$-times-offloaded request is not accepted by the fog node $j$ ($\delta_{jl}$), it will be offloaded for the $(l+1)$'st time to fog node $j$'s best neighbor ($\beta_{j(l+1)}$). $I_j$ hence could be also be expressed as $\delta_{j0}$;  type Light requests that are not offloaded so far. 

Finally, if a request is already offloaded $e_\mathcal{M}$ times, and is not accepted by fog node $j$, it will be offloaded to the cloud. This is realized by
\begin{equation}
C_j=(1-P_j)\cdot \delta_{je_\mathcal{M}}. 
\end{equation}

$\beta_{j1}$ could be calculated using Equation (\ref{offload-from-iot}), because we have all the components of the equation. Though, to attain $\beta_{j2}$, \ldots, $\beta_{je_\mathcal{M}}$, we need to have the equations for $\delta_{jl}$'s. 

Consider fog node $j$, and recall $\delta_{jl}$ represents the type Light requests that are offloaded for the $l$'th time from neighboring fog nodes to fog node $j$. Let $\hat{j}$ be one such neighbor. The chances that fog node $\hat{j}$'s best neighbor is fog node $j$ (and hence offloads the requests to fog node $j$) is roughly $\frac{1}{deg(\hat{j})}$, where $deg(\hat{j})$ is the number of neighbors of fog node $\hat{j}$. Therefore, $\delta_{jl}$ can be obtained by the considering the chances of receiving offloaded type Light requests from neighboring fog nodes to fog node $j$ as 
\begin{equation}
\delta_{jl}=\negspace\sum_{\hat{j}\in \textit{nghbr}(j)}{\negdoublespace [\beta_{\hat{j}l}\cdot \frac{1}{deg(\hat{j})}]}~:~~~~~ 1\leq l\leq e_\mathcal{M}~,
\end{equation}
where $\textit{nghbr}(j)$ is the set of neighboring fog nodes of fog node $j$. We now have all the components of the system of equations, so we can get all the $\beta_{jl}$'s, hence all the $\delta_{jl}$'s, and hence the $\lambda_j$. As mentioned before, $\lambda'_j$ could be derived similarly.
\subsection{Arrival Rates to Cloud Servers}
We need the equations for the arrival rate to cloud servers ($l_k$ and $l'_k$) to evaluate the average waiting time of cloud servers. The incoming traffic to the cloud servers are both from IoT nodes and fog nodes (refer to Fig. \ref{queueing-model-fog}). Similar to Equation (\ref{incoming-traffic-from-iot}), we can express the arrival rate of type Light requests to cloud server $k$ as
\begin{equation}\label{incoming-traffic-from-fog-iot}
l_k=\negspace\sum_{i\in g^{-1}(k)}{\negspace[\gamma^{}_i\cdot p_i^C]}+\negspace\sum_{j\in h^{-1}(k)}{\negspace C_j},
\end{equation}
where $g^{-1}(k)$ is a mapping function that indicates the set of IoT nodes that send their requests to cloud server $k$, and $h^{-1}(k)$ is a mapping function that indicates the set of fog nodes that offload requests to cloud server $k$. Equation for $l'_k$, the arrival rate of type Heavy requests to cloud server $k$, is derived similarly to Eq. (\ref{incoming-traffic-from-fog-iot}) by substituting $ \gamma'_i$ and $C'_j$. 
%IoT nodes are M/M/1 queues with arrival rate $\gamma^{}_i$ or $\gamma'_i$, therefore $I_j$ and $I'_j$ are Poisson (fraction of $\gamma^{}_i$ and $\gamma'_i$). $\beta_{j1}$ and $\beta'_{j1}$ are Poisson since they are fractions of $I_j$ and $I'_j$, respectively; hence $\delta_{j1}$ and $\delta'_{j1}$; hence all $\beta_{jl}$'s, $\beta'_{jl}$'s, $\delta_{jl}$'s, and $\delta'_{jl}$'s; hence $\lambda_j$ and $\lambda'_j$; $C_j$ and $C'_j$; and $l_k$ and $l'_k$.
\subsection{Waiting Time of Cloud Server} \label{delay_cloud}
We model the cloud server $k$ with $m_k$ internal processing units as $m_k$ M/G/1 queuing systems, with a load balancer that places the requests (uniformly) in the processing units. Thus, the total arrival rate to each M/G/1 queue is given by $\Lambda_k=\frac{l_k+l'_k}{m_k}$. Since we assume that the load balancer distributes the requests uniformly to the processing units, the average waiting time at a cloud server is equal to the average waiting time of one its processing units ($\overline{H}_k=\overline{\Delta}_k$). 

In order to obtain the average waiting time of a processing unit at the cloud server $k$, we use the Pollaczek-Khinchine formula to determine the average queue length, and use Little's law to obtain the average waiting time. Thus, the average waiting time of a processing unit at the cloud server $k$ will be: 
\begin{IEEEeqnarray}{rCl}\label{pk}
\overline{\Delta}_k=\frac{2\rho_k+\Lambda^2_k\sigma^2_k-\rho^2_k}{(2-2\rho_k)\Lambda_k},~~~~\rho_k=\Lambda_kE(\mathcal{S}_k).
\end{IEEEeqnarray}
where $E(\mathcal{S}_k)$ and $\sigma^2_k$ are the overall average and variance of service time of a request at a given processing unit of the cloud server $k$, respectively. At a processing unit of the cloud server $k$, the service time for a Light request, $s_k$ is exponentially distributed with an average service time of $\mathrm{Z}_k$, and the service time for a Heavy request, $s'_k$ is exponentially distributed with an average service time of $\mathrm{Z}'_k$. We can derive $E(\mathcal{S}_k)$ and $\sigma^2_k$ as:
\begin{IEEEeqnarray}{rCl}\label{pk}
E(\mathcal{S}_k)&=&(\frac{l_k}{l_k+l'_k})\cdot \mathrm{Z}_k + (\frac{l'_k}{l_k+l'_k})\cdot \mathrm{Z}'_k,\\
\sigma^2_k&=&E(\mathcal{S}^2_k)-E(\mathcal{S}_k)^2, ~~\textrm{where}\IEEEnonumber\\
E(\mathcal{S}^2_k)&=&(\frac{l_k}{l_k+l'_k})\cdot E(s^2_k)+(\frac{l'_k}{l_k+l'_k})\cdot E(s'^2_k).
\end{IEEEeqnarray}

%We can obtain $E(s^2_k)$ using the equation below and knowing that $s_k$ has an exponential  distribution with an average of $\mathrm{Z}_k$
%\begin{equation}
%	E(s^2_k)=\textrm{Var}(s_k)+E(s_k)^2=(\mathrm{Z}_k)^2+(\mathrm{Z}_k)^2=2(\mathrm{Z}_k)^2.
%\end{equation}
%Similarly, $E(s'^2_k)=2(\mathrm{Z}'_k)^2$.
\section{Numerical Evaluation} \label{results}
In this section we evaluate the proposed mechanism through simulation, and also compare the simulation results with our analytical model. We first discuss the simulation settings in the following subsection, and then explain the results in the next subsection.
\subsection{Simulation Settings}
\subsubsection{Network Topology}The network topology is a graph (hierarchical, similar to Fig. \ref{fig:arch}) with 500 IoT nodes, 25 fog nodes, and 6 cloud servers. The IoT node either processes its own request, or sends it to its corresponding fog neighbor or to one of the cloud servers. If the request is sent to the fog layer, based on the proposed scheme, the request could be offloaded to other fog nodes or to the cloud. The topology of the fog nodes in the fog layer is generated randomly in each experiment using a random graph generator with average node degree of 3. IoT nodes are associated with the fog node that has the smallest distance from them (i.e. has the smallest propagation delay).

\subsubsection{Link Bandwidth}If an IoT node generates type Light requests (e.g. sensor), the communication between the IoT node and its corresponding fog node is assumed to be through IEEE 802.15.4, or NB-IoT, or ZigBee, in which the transmission rates are 250 Kbps. If the IoT node generates Heavy requests (e.g. traffic camera), the communication between the IoT node and its corresponding fog node is assumed to be through IEEE 802.11 a/g, and the transmission rate is 54 Mbps. The link rates between fog nodes in one domain are 100 Mbps and the link rates on the path from fog nodes to cloud servers are 10 Gbps.

\subsubsection{Propagation and Transmission Delay}The propagation delay can be estimated by halving the round-trip time, $RTT$, which itself can be expressed as $RTT(\text{ms}) = 0.03\times \text{distance (km)}+5$ \cite{qureshi2010power}. Using this, we assume the propagation delay between the IoT nodes and the fog nodes, among fog nodes, and between fog nodes and the cloud servers are uniformly distributed between U[1,2], U[0.5,1.2], and U[15,35] respectively (in ms). Request lengths are exponentially distributed with an average length of 100 bytes for light processing tasks, and 80 KB for heavy processing tasks. We assume that the length of the response is the same as the length of its corresponding request, on average. 

\subsubsection{Processing Delay}To obtain realistic values for the processing ratio of IoT nodes to fog nodes, we looked at the processing capabilities of the Arduino Uno R3 microcontroller (an example of IoT node generating Light requests) and an Intel dual-core i7 CPU (an example of fog node). In the worst case, a fog node's processing capability is found to be around 3000 times faster than that of an IoT node generating type Light requests (``Fog-to-IoT-Light ratio''), and 200 times faster than that of an IoT node generating type Heavy requests (`Fog-to-IoT-Heavy ratio''). We also assume that a cloud server is 100 times faster than a fog node (``Cloud-to-Fog ratio''), on average, and that the average processing time of IoT node for Light and Heavy requests is 30 ms and 400 ms, respectively. Other simulation parameters are summarized in Table \ref{sim-parameters} for 5 different settings for parameters. To account for the variation of values of the above parameters in real IoT-fog-cloud applications, we altered the parameters uniformly as: Fog-to-IoT-Light ratio, U[500,4000]; Fog-to-IoT-Heavy ratio, U[100,400]; and Cloud-to-Fog ratio, U[50,200]; we found that the result (average delay) fluctuates only by -0.77\% to +5.51\%. 

\begin{table}[!t] 
	\caption{Simulation Parameters ($p^F_i$ values are shown for AFP)}
	\renewcommand{\arraystretch}{1}
	\label{sim-parameters}
	\centering
	{\scriptsize 
	\begin{tabular}{|c||cccccccc|}
		\hline
		Setting&$p^I_i$&$p^F_i$&$b_i$&$\theta_j$&$e_\mathcal{M}$&$\gamma^{}_i$&$\gamma'_i$&$q$\\
		\hline
		1 & 0 & 1 & 0.8 & 0.2 & 1 & 0.1 & 0.25 & - \\
		2 & 0 & 0.85 & 0.5 & 0.0002 & - & 0.5 & 0.6 & 0.5 \\
		3 & 0.1 & 0.75 & - & 0.2 & 1 & 0.05 & 0.005 & 0.5\\
		4 &-& - & 0.9 & 0.0002 & 1 & 0.01 & 0.001 & 0.5\\
		5 &0& 0.75 & 0.02 & 0.0002 & 1 & 0.1 & 0.05 & 0.5\\
		\hline
	\end{tabular}
}
\end{table}
\subsubsection{Operation Modes} To see the benefits of the proposed offloading scheme, we compare the average service delay in three modes of the IoT-fog-cloud operation. The proposed scheme is labeled as All Fog Processing (AFP), while other possible modes are labeled Light Fog Processing (LFP) and No Fog Processing (NFP). In No Fog Processing (NFP) mode, the IoT node either processes its own requests, or sends them directly to the cloud (that is, no request is sent to the fog). In this case, $p^F_i = 0$, and $p^C_i = 1 - p_i^I$ for both Light and Heavy request types. Conversely, the other two modes benefit from processing requests in the fog layer.

In All Fog Processing (AFP) mode, the IoT node either processes its own requests, or sends them to the fog or to the cloud; in AFP, both request types Light and Heavy can be sent to and processed in the fog layer. The values of $p^I_i$ and $p^F_i$ are as shown in Table \ref{sim-parameters}, and are the same for both type Light and type Heavy requests. Light Fog Processing (LFP) is similar to AFP in all the cases, except that only type Light requests could be sent to the fog layer, and type Heavy request must be sent to the cloud if they are not processed at the IoT nodes. Said differently, type Heavy requests could be either processed at the IoT node or in the cloud, but type Light requests could be processed at IoT, fog, or cloud.  In this case, the values of $p^I_i$ and $p^F_i$ for type Light requests are as shown in Table \ref{sim-parameters}; however, for Heavy requests, $p^I_i$ is as shown in Table \ref{sim-parameters}, but $p^F_i$ is set to 0. In all the cases, the value of $p^C_i$ is determined by $p^C_i=1-p^I_i-p^F_i$.

\subsubsection{Figure Settings}5 different parameters for the scheme settings are considered in the simulation results, and these settings are summarized in Table \ref{sim-parameters}. Each sample point in the graphs for simulation is obtained using 1 million requests using an event-driven simulation. For even more detailed analysis, the delay for type Light and Heavy requests is plotted separately for the three modes in the color figures (Namely, AFP$_\textrm{H}$ and AFP$_\textrm{L}$). All time units are in ms. In Fig. \ref{sim-delay-e_M} and \ref{sim-delay-b-total}, in addition to simulation values (black), the results of our analytical model are plotted (gold), to show the accuracy of the analytical model.

%\begin{table}[!t] 
%	\caption{Probability of Handling Requests in Different Layers}
%	\label{mode-probabilities}
%	\centering
%		\begin{tabular}{|c||cc|cc|cc|}
%			\hline
%			~&\multicolumn{2}{c|}{AFP}&\multicolumn{2}{c|}{LFP}&\multicolumn{2}{c|}{NFP}\\
%			Mode&Light&Heavy&Light&Heavy&Light&Heavy\\
%			\hline
%			$p^I_i$&$\geq0$&$\geq0$&$\geq0$&$\geq0$&$\geq0$&$\geq0$\\
%			$p^F_i$&$>0$&$>0$&$>0$&0&0&0\\		
%			$p^C_i$&$\geq0$&$\geq0$&$\geq0$&$\geq0$&$\geq0$&$\geq0$\\	
%			\hline
%		\end{tabular}
%\end{table}

\subsection{Numerical Results}
\begin{figure}[!t]
	\centering
	\subfloat[]{\includegraphics[width=0.485\linewidth]{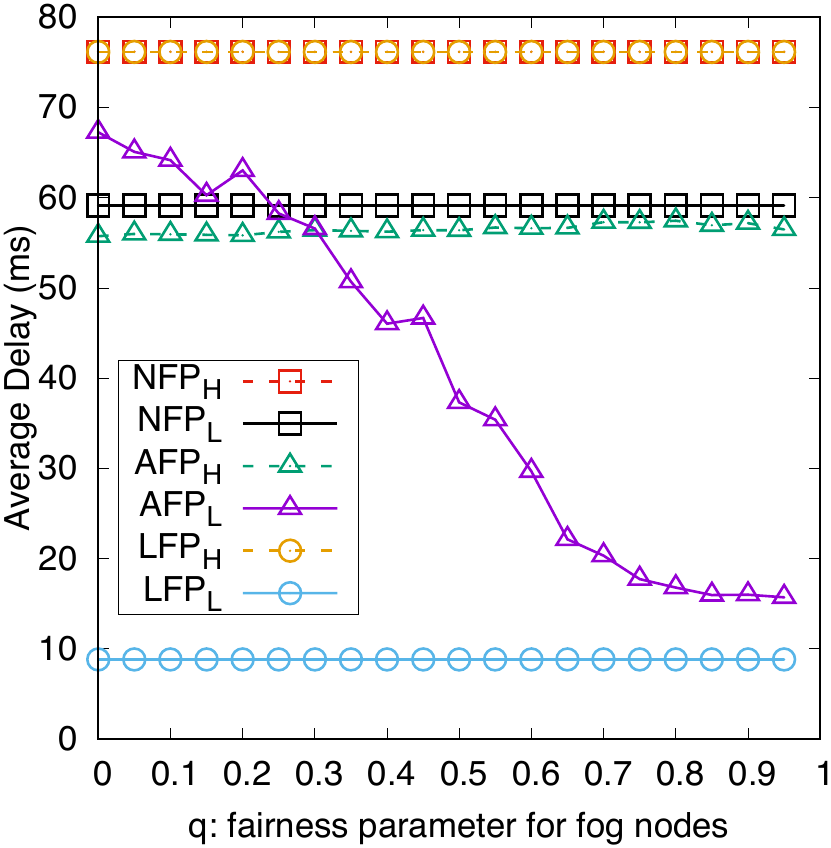}
		\label{sim-delay-q}}
	\subfloat[]{\includegraphics[width=0.515\linewidth]{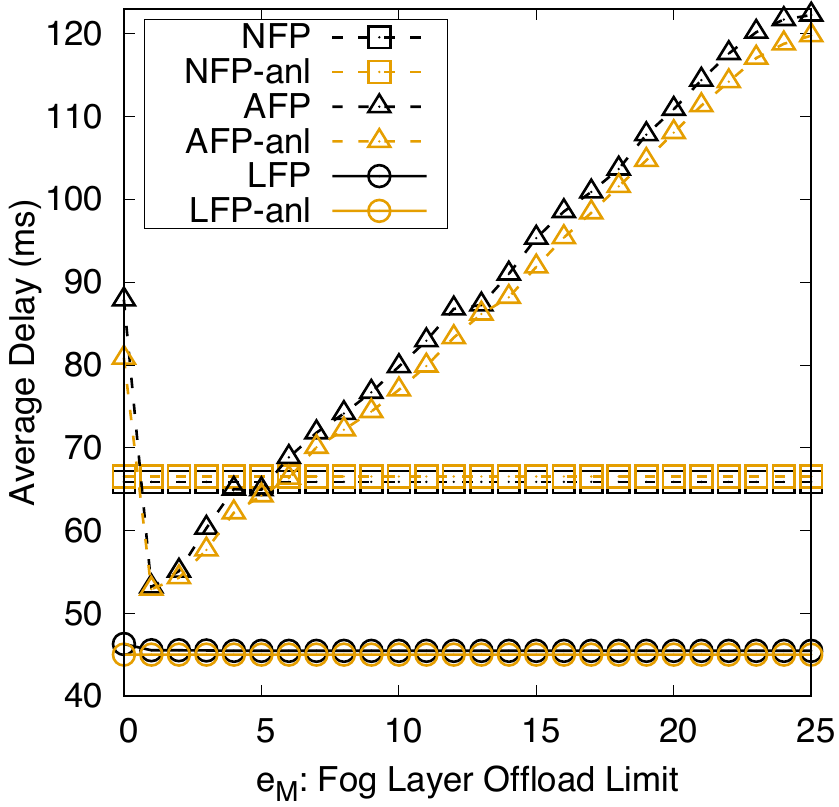}
		\label{sim-delay-e_M}}\\
	\subfloat[]{\includegraphics[width=0.5\linewidth]{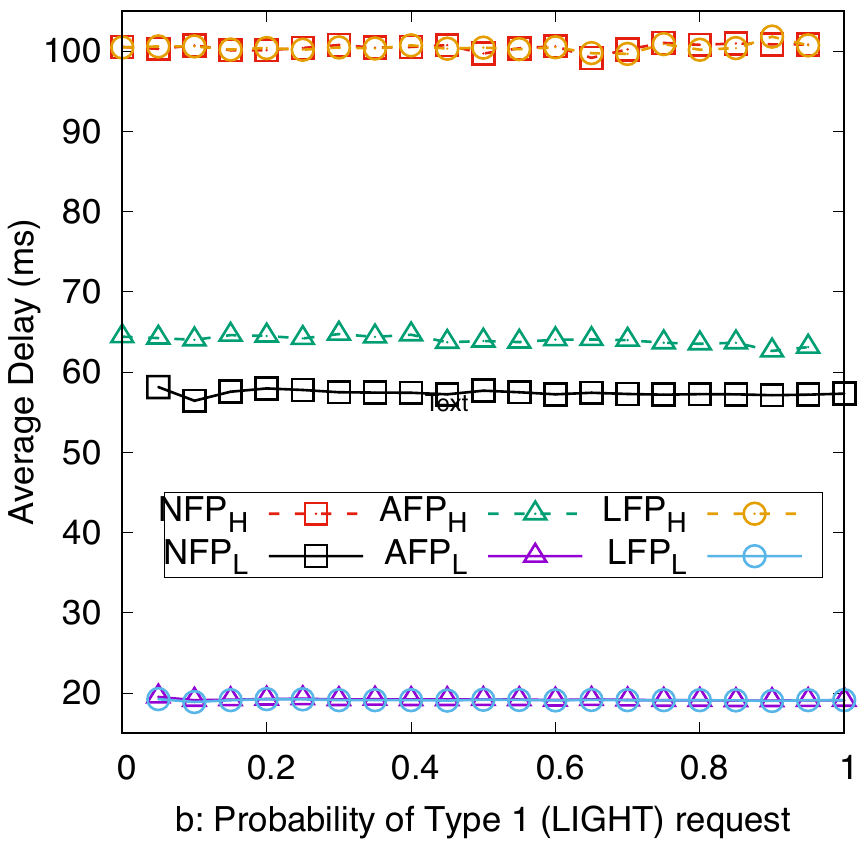}
		\label{sim-delay-b}}
	\subfloat[]{\includegraphics[width=0.5\linewidth]{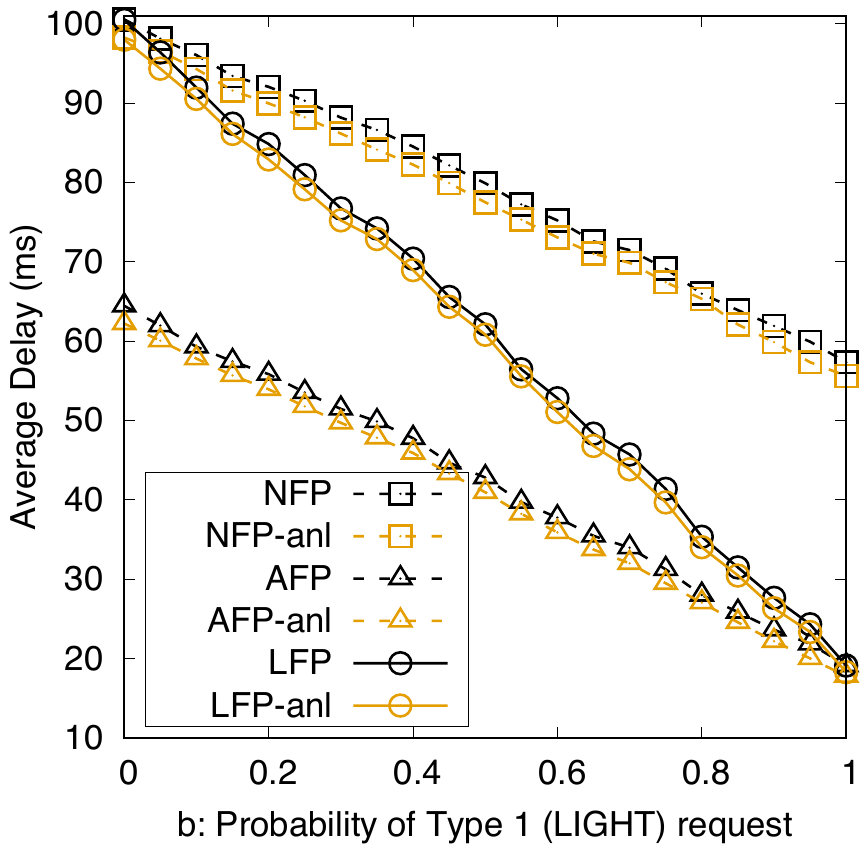}
		\label{sim-delay-b-total}}\\
	\subfloat[]{\includegraphics[width=0.487\linewidth]{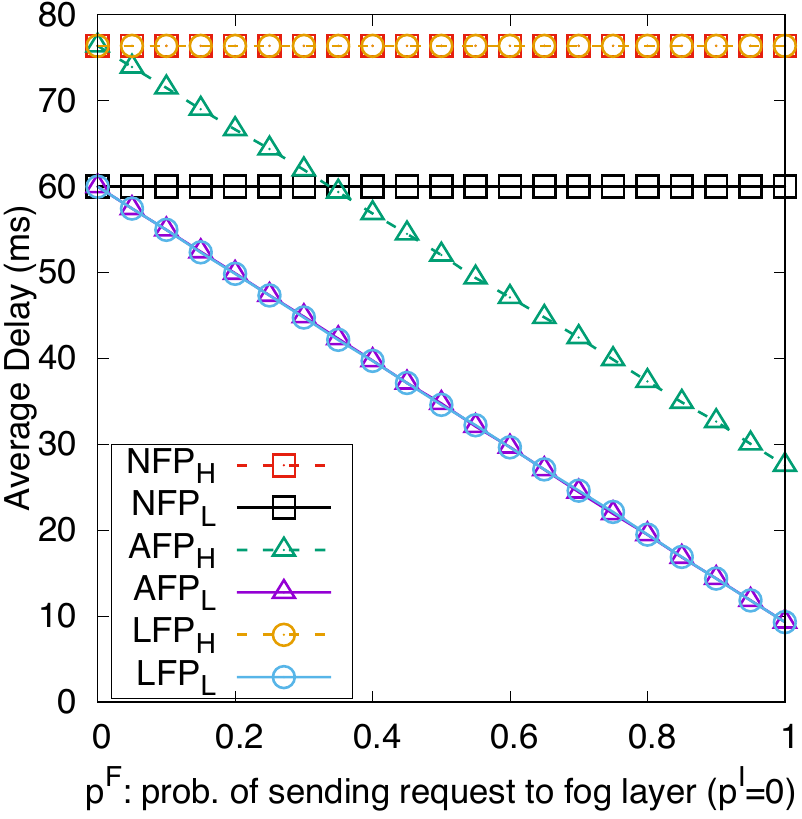}
		\label{sim-delay-P0}}
	\subfloat[]{\includegraphics[width=0.513\linewidth]{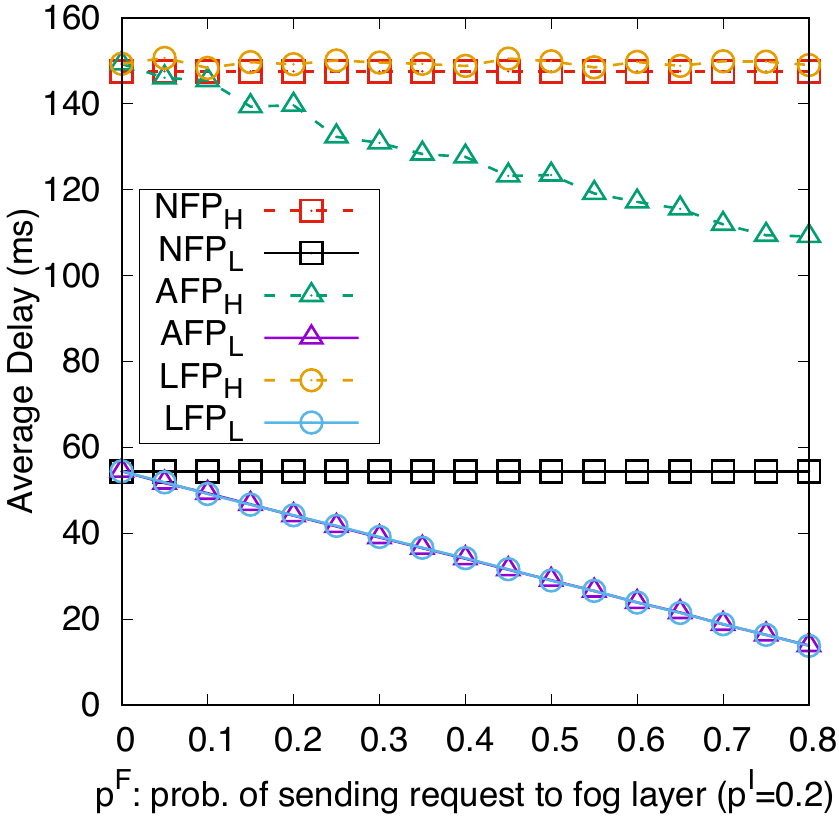}
		\label{sim-delay-P02}}

	\label{sim}
\end{figure}

Figure \ref{sim-delay-q} shows the average  delay as a function of fairness parameter $q$. For this figure, the simulation Setting 1 in Table \ref{sim-parameters} is used. Recall that $q$ is the fairness parameter, and when $q$ is closer to 1, more priority is given to light processing tasks. Thus when $q$ is closer to 1, the delay of light processing tasks is decreased and the delay of heavy processing tasks is increased. Note that this change is only seen in AFP, as the fairness parameter $q$ is not defined in NFP (there is no fog) and LFP (all Light requests).

Figure \ref{sim-delay-e_M} shows the average service delay as a function of $e_\mathcal{M}$ (obtained using simulation Setting 2). For AFP, the optimal value of $e_\mathcal{M}$ where the service delay is minimum is achieved for $e_\mathcal{M}=1$ using the parameters mentioned in simulation Setting 2, and when $e_\mathcal{M}>5$, AFP performs worse than NFP. It is interesting to see that changes in $e_\mathcal{M}$ do not change the average service delay in LFP, since the incurred transmission and propagation delay to offload a request among fog nodes is negligible for Light requests with small length. The suffix ``-anl'' in the legend of the figures denotes the analytical model values for the corresponding mode of the policy. It can be inferred that the analytical model results closely match with the simulation values.

Figure \ref{sim-delay-b} and \ref{sim-delay-b-total} (obtained using simulation Setting 3) show the average service delay as a function of $b_i$, the probability that a generated request at IoT node $i$ is type Light. Figure \ref{sim-delay-b} shows that the average service delay for both Heavy and Light requests do not change notably when the percentage of Light and Heavy requests change. This is because we are looking at each of the task types separately, and this is oblivious to the effect of $b_i$. By comparing the delay of Light and Heavy processing tasks in Fig. \ref{sim-delay-b} for the three modes, it is clear that the AFP has the lowest average delay. Figure \ref{sim-delay-b-total} illustrates the interesting relationship between average service delay (of combined Light and Heavy requests) in the three modes, while $b_i$ changes. It can be seen that AFP, in general, outperforms LFP and NFP in terms of average service delay; however, when the percentage of Light requests in the network increases, LFP's performance gets closer to that of AFP. This is due to having fewer heavy processing tasks in the network that make the average service delay larger. 

Figure \ref{sim-delay-P0} and \ref{sim-delay-P02} (obtained using simulation Setting 4) show the effects of $p^I_i$, $p^F_i$, and $p^C_i$ on the average service delay. Both figures show how the average service delay is reduced under each policy when the probability of sending requests to fog nodes increases. In Fig. \ref{sim-delay-P0}, $\forall i:p^I_i=0$ and it is clear that the performance of both LFP and AFP is better than that of NFP, as the delays in all the cases are lower. For Fig. \ref{sim-delay-P02}, $\forall i:p^I_i=0.2$ and it can be seen that the overall delay has been increased, due to the weak processing capabilities of IoT nodes. Yet, the overall delay is decreased to from 60 ms to 18 ms for light processing tasks, and from 150 ms to 117 ms, for heavy processing tasks. In this figure, it is also evident that the performance of LFP and AFP is better than that of NFP.

\begin{figure}[!t]
	\centering
	\subfloat[]{\includegraphics[width=0.5\linewidth]{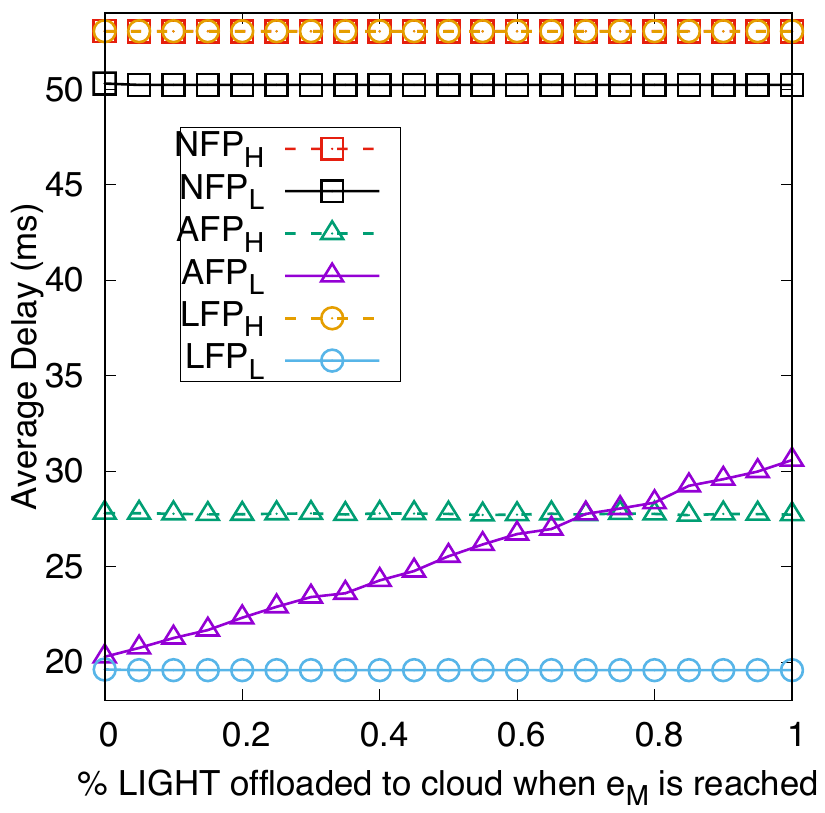}
		\label{sim-delay-perc}}
	\subfloat[]{\includegraphics[width=0.5\linewidth]{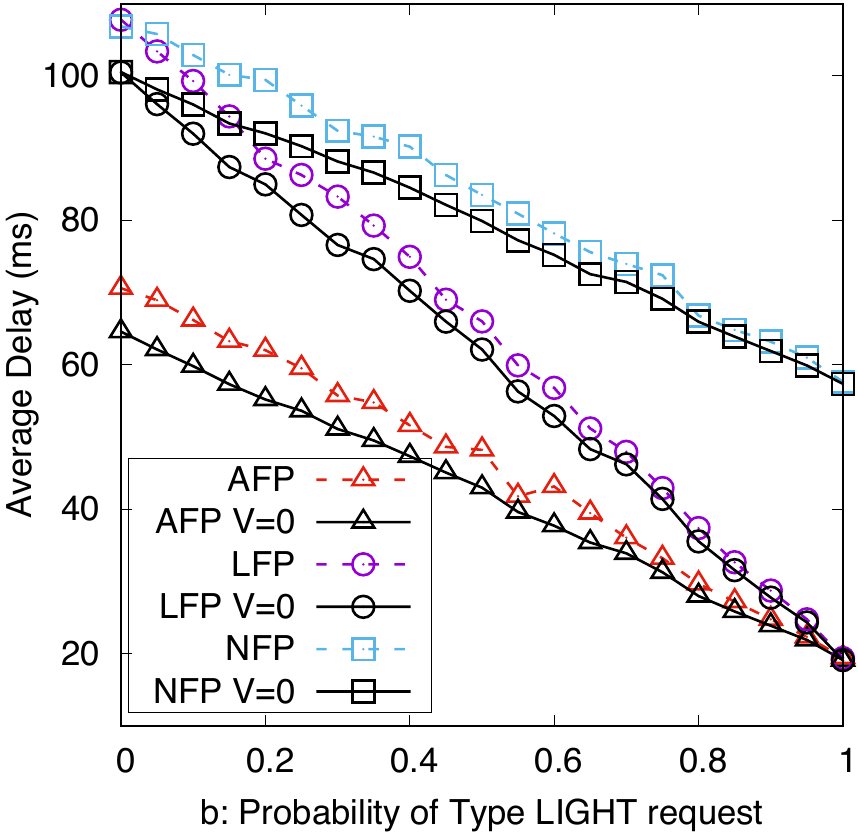}
		\label{sim-delay-gamma-var}}\\
	\subfloat[]{\includegraphics[width=0.64\linewidth]{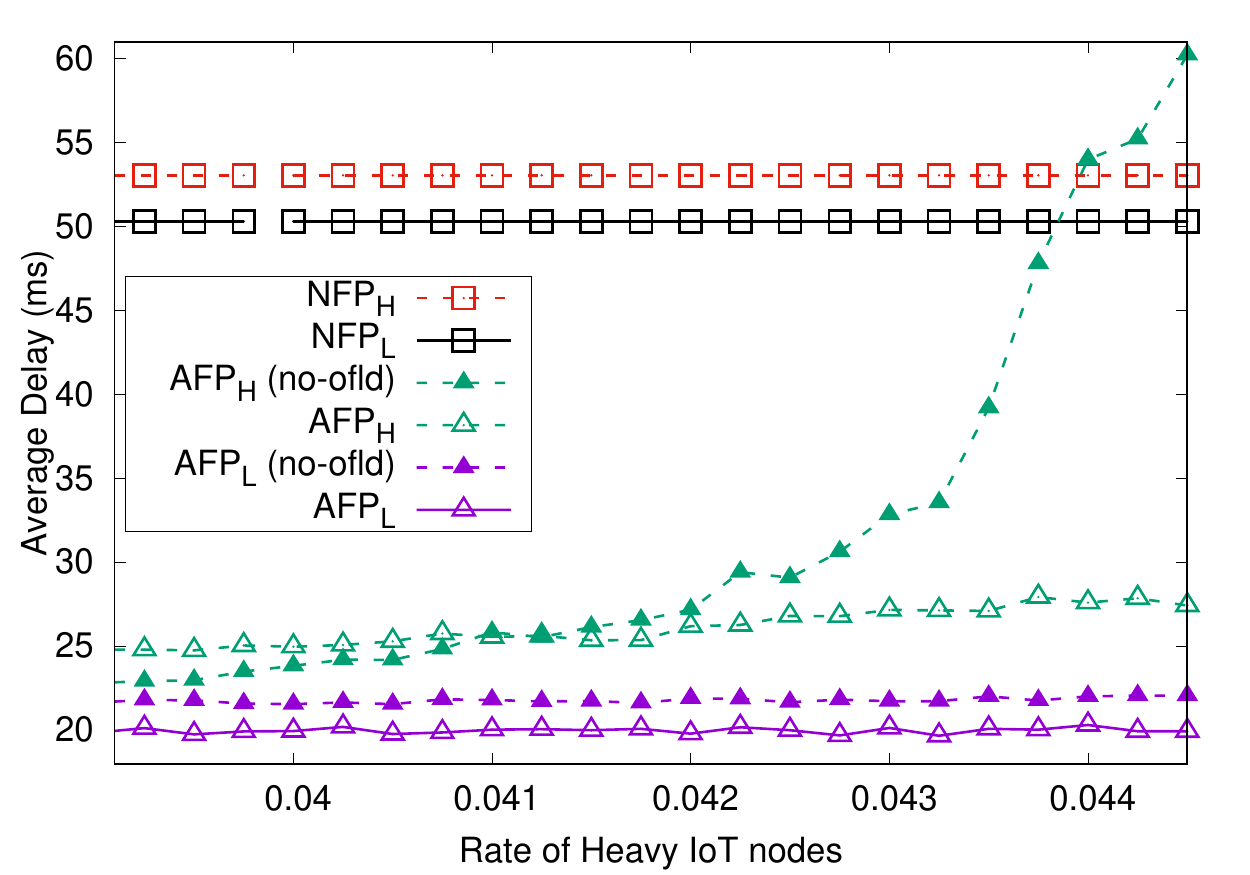}
		\label{sim-delay-off-nooff}}
	\subfloat[]{\includegraphics[width=0.39\linewidth]{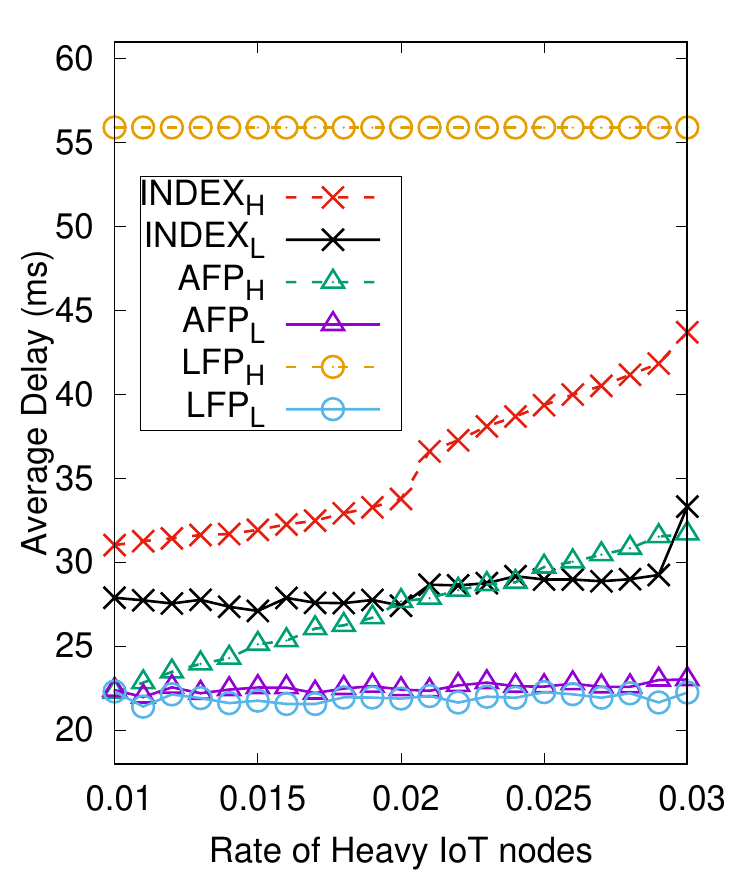}
		\label{sim-delay-index}}
	\caption{Comparison of variations in the proposed policy and other policies.} 
	\label{sim-var-policy}
	
\end{figure}

%\subsubsection{Variations of Policy}
In the rest of this section, we study the results of some modification to the proposed offloading policy. Figure \ref{sim-delay-perc} is shown to understand the benefits of offloading Heavy and Light requests to the cloud when the number of offloads of requests reaches $e_\mathcal{M}$ (simulation Setting 5). In this modification, when the number of offloads reaches $e_\mathcal{M}$, Heavy requests are always offloaded to the cloud, whereas only a certain percentage of type Light requests are offloaded to the cloud. This percentage is represented on the x-axis of Fig. \ref{sim-delay-perc}. It can be seen that if type Light requests are not offloaded to the cloud when the number of offloads reaches $e_\mathcal{M}$, service delay is minimum. This is due to the large propagation delays from the fog nodes to the cloud for type Light requests, relative to their small processing delays. This figure suggests that it is better to accept the type Light requests at a fog node, instead of sending them to the cloud, when $e_\mathcal{M}$ is reached. Note that in this figure, service delay does not change noticeably for LFP. This is because the value of $\theta_j$ is big enough for Light requests, that offloading does not happen much.

Figure \ref{sim-delay-gamma-var} (simulation Setting 3) shows the effects of variance in the rate of generating requests from IoT nodes ($\gamma^{}_i$ and $\gamma'_i$). The black points in the figure (with suffix ``V=0'') are obtained when there is no variance in the rate of generating requests, and the color points are obtained when there is some variance in the rate of generating requests. For the color points, the rate of generating requests is a normal distribution with standard deviation equal to the average rate of generating requests. In both cases, the average rates of generating requests are $\gamma^{}_i=0.05$ and $\gamma'_i=0.005$, as reported in Table \ref{sim-parameters} for simulation Setting 3. It is perceived that when there is variance in the rates of generating requests, service delay gets larger. This increase is aggravated when there are more type Heavy requests in the network, because when there are more type Heavy requests, the effect of variance in rates of generating requests is more notable. As seen before, among all modes, AFP has the best performance, and LFP's performance gets closer to that of AFP, when the percentage of Light requests in the network increases.

Figure \ref{sim-delay-off-nooff} (simulation Setting 5) is plotted to compare the benefits of processing requests in the fog layer (fog, in general), to the benefits of offloading requests in the fog layer (proposed offloading policy). Similar to some of the previous figures, service delay for type Light and Heavy are plotted separately in these figures. The x-axis represents $\gamma'_i$, the rate of generating Heavy requests from IoT nodes. The suffix ``(no-ofld)'' in the legend of the figure denotes the mode when offloading of fog nodes is disabled. In other words, in ``(no-ofld)'' if a request is sent to a particular fog node, the fog node always accepts it regardless of how busy it is. This mode denotes the general fog node when the proposed offloading policy is not used. In Fig. \ref{sim-delay-off-nooff}, this mode is compared to the mode when our proposed offloading policy is used.

It is clear that the average service delay for AFP type Light requests (colored purple) is minimized when the proposed offloading policy is used. On the contrary, for AFP type Heavy requests (colored green), the average service delay in the proposed offloading policy is not always less than that of the ``(no-ofld)''. When the arrival rate of type Heavy requests in the network is high, the proposed offloading policy greatly helps to reduce the average service delay for type Heavy requests. Yet, when the arrival rate of type Heavy requests is not high (in this setting when $\gamma'_i<0.041$), the proposed offloading policy has slightly higher average service delay for type Heavy requests in comparison to ``(no-ofld)''. So when arrival rate of type Heavy requests in a network is high, it is beneficial to use the proposed offloading policy.

In Fig. \ref{sim-delay-index} (simulation Setting 5) we compare the performance of our scheme with that of the scheme proposed in \cite{indexbased}, referred to as ``index policy''. In the index policy, for task assignment, edge clouds (cloud servers deployed at the edge of the network) calculate a number (called index) based on their queueing status, and broadcast this number to the mobile subscribers. Mobile subscribers will then select an edge cloud with the smallest index, and send their request to it. We can consider mobile subscribers as IoT nodes and the edge clouds as fog nodes and compare the performance our scheme with that of the index policy. Extra parameters of the index policy are set as: $\xi_n=1$ and $\eta=0.2$.

The delay for type Light and Heavy requests in the index policy is denoted in Fig. \ref{sim-delay-index} by ``INDEX$_\textrm{L}$'' and ``INDEX$_\textrm{H}$'', respectively. We can see that our scheme (AFP) performs better than the index policy. The delay of type Light and Heavy in AFP are significantly less that the delay of type Light and Heavy in the index policy. This is because the propagation delay and transmission rate between IoT nodes and fog nodes are not considered in the index policy; the index is simply calculated based on the fog node's queue status only. The LFP mode is drawn in this figure as an upper bound for the delay of type Heavy requests. As seen in other figures, the delay of type Light requests in AFP and LFP are very close.

\section{Conclusion} \label{conclusion}
The vision of fog computing is studied in this paper as a complement to cloud computing and an essential ingredient of the IoT. We introduced a framework for handling IoT request in the fog layer and an analytical model to formulate service delay in the IoT-fog-cloud scenarios. We showed how our delay-minimizing fog offloading policy can be beneficial for the IoT. Various numerical results are included to support our claims by showing how changes in parameters could affect the average service delay, and to show how our analytical model can be used to describe the performance of the policy.

Our analytical model can support other fog computing policies. For example, when the decision to offload a task is not based on queueing status, one can replace $P_j$ and $L_{ij}$, with the desired equations based on their policy. As future work, one can consider additional dimensions of IoT requests, such as the amount of data that the request carries. Additionally, one can propose an approach to adjust fog nodes' threshold ($\theta_j$'s) dynamically. Moreover, it may be interesting to investigate delay, cost, and energy tradeoffs in fog offloading schemes.




%\bibliographystyle{IEEEtran}

\bibliography{Master}{}
\bibliographystyle{ieeetr}

\vfill


\end{document}